\documentclass[aps,preprint,showpacs,superscriptaddress,groupedaddress]{revtex4}  % for double-spaced preprint
\usepackage{graphicx}  % needed for figures
\usepackage{float}
\usepackage{dcolumn}   % needed for some tables
\usepackage{bm}        % for math
\usepackage{amssymb}   % for math
\usepackage{slashed}   % for Dirac Slash
\usepackage{amsmath}   % for mutiline eqn
\usepackage{simplewick} % for contraction
\usepackage{verbatim}  % for multi-line comment
\usepackage{color} % for textcolor
\usepackage{appendix} % for appendix
\begin{document}

% The following information is for internal review, please remove them for submission
\widetext

% the following line is for submission, including submission to the arXiv!!
% \hspace{5.2in} \mbox{Fermilab-Pub-04/xxx-E}

\title{Factorization of Radiative Leptonic Decays of $B^-$ and $D^-$ Mesons}
\affiliation{School of Physics, Nankai University, Tianjin 300071, P.R. China}
\author{Ji-Chong Yang} \affiliation{School of Physics, Nankai University, Tianjin 300071, P.R. China}
\author{Mao-Zhi Yang} \affiliation{School of Physics, Nankai University, Tianjin 300071, P.R. China}
%
% visitor_addresses.tex                        3 December 2011
%  available symbols are:
%  $\ast, \dag, \ddag, \S, \P, $\|$, $\ast\ast$, \dag\dag, \ddag\ddag ,\#
%
%
\vskip 0.25cm
% D0 authors (remove the first 3 lines
                             % of this file prior to submission, they
                             % contain a time stamp for the authorlist)
                             % (includes institutions and visitors)
\date{\today}

\begin{abstract}
In this work, we study the factorization of the radiative leptonic decays of $B^-$ and $D^-$ mesons, the contributions of the order $O(\Lambda _{\rm QCD}\left/m_Q\right.)$ are taken into account. The factorization is proved to be valid explicitly at the order $O(\alpha _s\Lambda _{\rm QCD}\left/m_Q\right.)$. The hard kernel is obtained. The numerical results are calculated using the wave-function obtained in relativistic potential model. The $O(\Lambda _{\rm QCD}\left/m_Q\right.)$ contribution is found to be very important, the correction to the decay amplitudes of $B^-\to \gamma e\bar{\nu}$ is about $20\% - 30\%$. For $D$ mesons, the $O(\Lambda _{\rm QCD}\left/m_Q\right.)$ contributions are more important.
\end{abstract}

\maketitle

\section{\label{sec:level1}Introduction}

The study of the heavy meson decays is an important field in high energy physics. In recent years, both experimental and theoretical studies have been improved greatly \cite{experimatal data,theoretical development1,theoretical development2}. However, the limitation in understanding and controlling the non-perturbative effects in strong interaction is so far still a problem. Varies theoretical methods on how to deal with the non-perturbative effects have been developed. An important approach is to separate the hard and soft physics which is known as factorization \cite{DIS and Dran-Yan,Collins method}. This method has been greatly developed in recent years \cite{Sachrajda other work}. The idea of factorization is to absorb the infrared (IR) behaviour into the wave-function, the matrix element can be written as the convolution of wave-function and hard kernel
\begin{equation}
\begin{split}
&F=\int dk \Phi(k)T_{\rm hard}(k)\\
\end{split}
\label{eq.0.1}
\end{equation}
The wave-function should be determined by non-perturbative methods.

The radiative leptonic decay of heavy mesons provides a good opportunity to study the factorization approach, where strong interaction is involved only in one hadronic external state. Except for that, with a photon emitted out, more details about the wave-function of the hadronic bound state can be exploited. Many works has been done on the factorization of this decay mode. In Ref.~\cite{TM Yan work}, the 1-loop QCD correction is calculated in the large energy effective theory, and they found the factorization will depend on the transverse momentum. In Ref.~\cite{Genon and Sachrajda Radivative Leptonic} and \cite{Neubert work}, factorization is proved in leading order of $1\left/ m_Q\right.$ expansion in the frame of QCD factorization \cite{DIS and Dran-Yan,Collins method}, where the heavy quark is treated in the heavy quark effect theory (HQET) \cite{theoretical development2,heavy quark expansion}. In Ref.~\cite{SCET work neubert,Daniel Wyler work}, the factorization is constructed using the soft-collinear effective theory (SCET) \cite{SCET work 1,SCET work 2}.

In this work, factorization in the radiative leptonic decays of heavy mesons is revisited. The work of Ref.~\cite{Genon and Sachrajda Radivative Leptonic,Neubert work} is extended by taking into account the contributions of the order of $O(\Lambda _{\rm QCD}\left/m_Q\right.)$. The factorization is proved to be still valid explicitly. We also find the factorization is valid at any order of $O(\Lambda _{\rm QCD}\left/m_Q\right.)$. The numerical results shows that, the $O(\Lambda _{\rm QCD}\left/m_Q\right.)$ correction is very important for the $B$ and $D$ mesons, the correction can be as large as $20\%-30\%$.

The remainder of the paper is organized as follows. In Sec.~\ref{sec:level2}, we discuss the kinematic of the radiative decay and the wave-function. In Sec.~\ref{sec:level3}, we present the factorization at tree level. In Sec.~\ref{sec:level4}, the 1-loop corrections of the wave-function are discussed. The factorization at 1-loop order is presented in Sec.~\ref{sec:level5}. In Sec.~\ref{sec:level6}, we briefly discuss the resummation of the large logarithms. The numerical results are presented in Sec.~\ref{sec:level7}. And Sec.~\ref{sec:level8} is a summary.

\section{\label{sec:level2}The kinematic}

\begin{figure}
\includegraphics[scale=0.75]{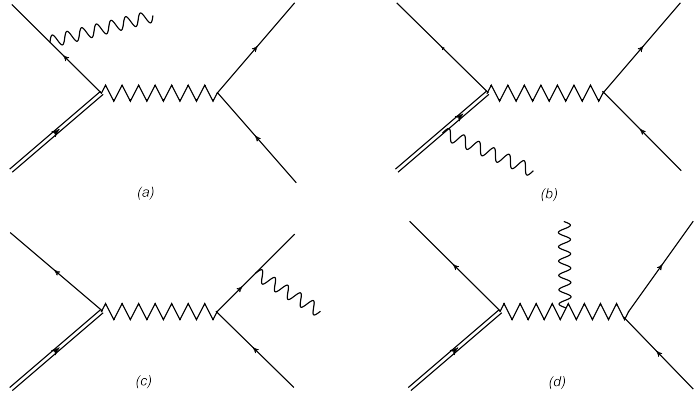}
\caption{\label{Fig1:epsart} tree level amplitudes, the double line represents the heavy quark propagator but not the HQET propagator.}
\end{figure}

The $B$ or $D$ meson is constituted with a quark and an anti-quark, where one of the quarks is a heavy quark, and the other is a light quark. The Feynman diagrams at tree level of the radiative leptonic decay can be shown as Fig.~\ref{Fig1:epsart}. The contribution of Fig.~\ref{Fig1:epsart}.d.\ is suppressed by a factor of $1\left/M_w{}^2\right.$, and can be neglected. The amplitudes of Fig.~\ref{Fig1:epsart}.a, b and c can be written as
\begin{equation}
\begin{split}
&\mathcal{A}_a^{(0)}=\frac{- i e_q G_F V_{Qq}}{\sqrt{2}}\bar{q}(p_{\bar{q}})\slashed \varepsilon _{\gamma }^*\frac{\slashed p_{\gamma }-\slashed p_q}{2p_{\gamma}\cdot p_{\bar{q}}}P_L^{\mu } Q(p_Q)\left(lP_{L\mu}\bar{\nu}\right)\\
&\mathcal{A}_b^{(0)}=\frac{- i e_Q G_F V_{Qq}}{\sqrt{2}}\bar{q}(p_{\bar{q}})P_L^{\mu }\frac{\slashed p_Q-\slashed p_{\gamma}+m_Q}{2p_Q\cdot p_{\gamma}}\slashed \varepsilon _{\gamma }^* Q(p_Q)\left(lP_{L\mu}\bar{\nu}\right)\\
&\mathcal{A}_c^{(0)}=\frac{- e G_F V_{Qq}}{\sqrt{2}}\bar{q}(p_{\bar{q}})P_L^{\mu } Q(p_Q)\left(l\slashed \varepsilon _{\gamma }^*\frac{i(\slashed p_{\gamma }+\slashed p_l+m_l)}{2\left(p_{\gamma }{\cdot}p_l\right)}P_{L\mu}\bar{\nu}\right)
\end{split}
\label{eq.1.1}
\end{equation}
where $p_{\bar{q}}$ and $p_Q$ are the momenta of the anti-quark $\bar{q}$ and quark $Q$, respectively, $p_{\gamma}$, $p_l$ and $p_{\nu}$ are the momenta of photon, lepton and neutrino, $\varepsilon _{\gamma}$ denotes the polarization vector of photon, and $P_L^{\nu}$ is defined as $\gamma^{\mu}(1-\gamma _5)$.

We work in the rest-frame of the meson, and we choose the frame such that the direction of the photon momentum is on the opposite z axis, so the momentum of the photon can be written as $p_{\gamma}=(E_{\gamma}, 0, 0, -E_{\gamma})$, with $0\le E_{\gamma} \le m_Q \left/ 2 \right.$.

To study the factorization, we consider the state of two free quark and anti-quark at first. The wave-function of the two quark and anti-quark state is defined as
\begin{equation}
\Phi (k_q,k_Q) = \int d^4xd^4y \exp (i k_q\cdot x)\exp (i k_Q\cdot y) <0|\bar{v}_{\bar{q}}(x)\left[x, y\right]u_Q(y)|\bar{q}Q>
\label{eq.1.2}
\end{equation}
where $\left[x, y\right]$ denotes the Wilson line \cite{Wilson line}. And the matrix element is defined as
\begin{equation}
F= <\gamma|\bar{v}_{\bar{q}}(x)P_L^{\mu}u_Q(y)|\bar{q}Q>
\label{eq.1.3}
\end{equation}
The prove of factorization is to prove that, up to 1-loop order, the matrix element can be written as the convolution of the wave-function $\Phi$ and a hard-scattering kernel $T$, where $T$ is IR finite and independent of the external state.

\section{\label{sec:level3}Tree level factorization}

We start with the matrix elements at tree level. Using the definition of the wave-function in coordinate space
\begin{equation}
\Phi_{\alpha\beta}(x, y)=<0|\bar{q}_{\alpha}(x)[x,y]Q_{\beta}(y)|\bar{q}^S(p_{\bar{q}}),Q^s(p_Q)>
\label{eq.2.2}
\end{equation}
where $S$ and $s$ are spin labels of $\bar{q}$ and $Q$, respectively. We find
\begin{equation}
\begin{split}
&\Phi^{(0)}_{\alpha\beta}(k_{\bar{q}}, k_Q)=(2\pi)^4\delta^4(k_{\bar{q}}-p_{\bar{q}})(2\pi)^4\delta^4(k_Q-p_Q) \bar{v}_{\alpha}(p_{\bar{q}})u_{\beta}(p_Q)
\end{split}
\label{eq.2.3}
\end{equation}
And then the matrix element can be written as
\begin{equation}
\begin{split}
&F^{(0)}=\int \frac{d^4k_{\bar{q}}}{(2\pi) ^4}\frac{d^4k_Q}{(2\pi)^4}\Phi^{(0)}(k_{\bar{q}}, k_Q)T^{(0)}(k_{\bar{q}}, k_Q)=\Phi^{(0)}\otimes T^{(0)}\\
\end{split}
\label{eq.2.4}
\end{equation}
With Eqs.~(\ref{eq.1.1}) and (\ref{eq.2.4}), we obtain the hard scattering kernel at tree level as
\begin{equation}
\begin{split}
&T_a^{(0)}=-e_q\frac{\slashed \varepsilon _{\gamma }^*\slashed p_{\gamma }-2 \varepsilon _{\gamma }^*\cdot k_q}{2p_{\gamma}\cdot k_{\bar{q}}}P_L^{\mu },\;\;\;\;\;T_b^{(0)}=-e_QP_L^{\mu }\frac{-\slashed p_{\gamma}\slashed \varepsilon _{\gamma }^*}{2k_Q\cdot p_{\gamma}}
\end{split}
\label{eq.2.5}
\end{equation}
In the expressions above, we have already assumed to consider the kinematical region $E_{\gamma}\sim m_Q$. The polarization vector of the photon does not have 0-component, as a result $(\slashed k_Q + m_Q)\slashed \varepsilon u_Q \left/ 2p_{\gamma}\cdot p_Q\right.$ is an order of $O(\Lambda _{\rm QCD}^2\left/ m_Q^2\right.)$ contribution and is neglected. The remaining terms of $T_b^{(0)}$, and the transverse part of $T_a^{(0)}$, which is $2 e_q \varepsilon \cdot k_{\bar{q}}P_L^{\mu}\left/2p_{\gamma}\cdot k_{\bar{q}}\right.$ are order of $O(\Lambda _{\rm QCD}\left/ m_Q\right.)$ contributions.

Exchange the Lorentz index in $\mathcal{A}_c$, we obtain
\begin{equation}
\begin{split}
&\mathcal{A}_c^{(0)}=\frac{- i e G_F V_{Qq}}{\sqrt{2}}\bar{q}(p_{\bar{q}})P_L^{\mu }\frac{(\slashed p_{\gamma }\slashed \varepsilon _{\gamma }^*+2\varepsilon _{\gamma }^*\cdot (p_Q+p_{\bar{q}}-p_{\nu}))}{2\left(p_{\gamma }{\cdot}(p_Q+p_{\bar{q}}-p_{\nu})\right)}Q(p_Q)\left(lP_{L\mu}\bar{\nu}\right)
\end{split}
\label{eq.2.9}
\end{equation}
We find
\begin{equation}
\begin{split}
&F_c^{(0)}=-e\bar{v}_{\bar{q}} P_L^{\mu }\frac{\slashed p_{\gamma }\slashed \varepsilon _{\gamma }^*+2\varepsilon \cdot (p_Q+p_{\bar{q}}-p_{\nu})}{2p_{\gamma }\cdot (p_Q+p_{\bar{q}}-p_{\nu})}u_Q,\;\;\;\;\;T_c^{(0)}=-eP_L^{\mu }\frac{\slashed p_{\gamma }\slashed \varepsilon _{\gamma }^*+2\varepsilon \cdot (k_Q+k_{\bar{q}}-p_{\nu})}{2p_{\gamma }\cdot (k_Q+k_{\bar{q}}-p_{\nu})}
\end{split}
\label{eq.2.10}
\end{equation}

This term is also an order of $O(\Lambda _{\rm QCD}\left/m_Q\right.)$ contribution.

\section{\label{sec:level4}1-loop correction of wave-function}

The expansion of the decay amplitude can be written as \cite{Genon and Sachrajda Radivative Leptonic}
\begin{equation}
\begin{split}
F&=F^{(0)}+F^{(1)}+\ldots =\Phi ^{(0)}\otimes T^{(0)}+\Phi ^{(1)}\otimes T^{(0)}+\Phi ^{(0)}\otimes T^{(1)}+\ldots
\end{split}
\label{eq.3.1}
\end{equation}

At the 1-loop level, the amplitude can be written as
\begin{equation}
F^{(1)}=\Phi ^{(1)}\otimes T^{(0)}+\Phi ^{(0)}\otimes T^{(1)}
\label{eq.3.2}
\end{equation}

\begin{figure}
\includegraphics[scale=0.9]{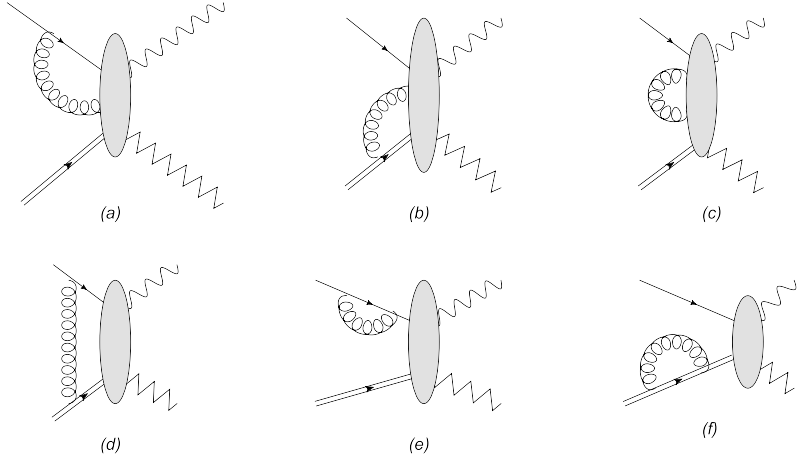}
\caption{\label{Fig2:epsart} The 1-loop correction of wave-functions. $\Phi^{(1)} \otimes T_a^{(0)}$ and $\Phi^{(1)} \otimes T_b^{(0)}$ are established in this figure.}
\end{figure}

The 1-loop corrections of $\Phi \otimes T$ come from the QCD interaction and the Wilson-Line. The later can be written as \cite{Genon and Sachrajda Radivative Leptonic,Wilson line}
\begin{equation}
\begin{split}
[x,y]&=\exp \left[ig_s\int _y^x d^4z z_{\mu} A^{\mu}(z)\right]=\sum _{\substack{n}}\frac{(ig_s)^n}{n!}\prod _{\substack{i}}^n\int _y^x d^4z_i z_{i\mu} A^{\mu}(z_i)
\label{eq.3.3}
\end{split}
\end{equation}
The corrections are shown in Fig.~\ref{Fig2:epsart}.

We use $\Phi _q^{(1)}$ to represent the correction with the gluon from the Wilson Line connected to the light quark external leg. So the correction in Fig.~\ref{Fig2:epsart}.a.\ can be written as
\begin{equation}
\begin{split}
&\Phi_q^{(1)}(k_{\bar{q}}, k_Q)=\int d^4x \int d^4y e^{ik_{\bar{q}}\cdot x}e^{ik_Q\cdot y}<0|\bar{q}_{\bar{q}}(x)ig_s\int _y^x dz z_{\mu}A^{\mu}(z) Q(y)\\
&\times ig_s\int d^4x_2 \bar{q} _{\bar{q}}(x_2)\slashed {A}(x_2)q _{\bar{q}}(x_2)|\bar{q}^S(p_{\bar{q}}),Q^s(p_Q)>
\end{split}
\label{eq.3.4}
\end{equation}
After the integration, the result is
\begin{equation}
\begin{split}
&\Phi_q ^{(1)}\otimes T^{(0)}=ig_s^2C_F\int \frac{d^d l}{(2\pi)^d} \frac{1}{l^2}\bar{v}_{\bar{q}}\gamma ^{\rho}\frac{(\slashed {l}+\slashed {p}_{\bar{q}} - m_{\bar{q}})}{(l+p_{\bar{q}})^2-m_{\bar{q}}^2} \int _0^1 d\alpha\left.\left(\frac{\partial T^{(0)}}{\partial k_q^{\rho}}-\frac{\partial T^{(0)}}{\partial k_Q^{\rho}}\right)\right| _{k_q=k',k_Q=K'} u_Q\\
&k'=p_{\bar{q}}+\alpha l,\;\;K'=p_Q-\alpha l\\
\end{split}
\label{eq.3.12}
\end{equation}
The procedure of the integration can be found in Appendix A.

Similar to $\Phi _q$, the correction in Fig.~\ref{Fig2:epsart}.b.\ can be written as
\begin{equation}
\begin{split}
&\Phi_Q ^{(1)}\otimes T^{(0)}=-iC_Fg_s^2\int _0^1 d\alpha \int \frac{d^d l}{(2\pi)^d}\frac{1}{l^2}
\bar{v}_{\bar{q}}\left.\left(\frac{\partial T^{(0)}}{\partial k_q^{\rho}}-\frac{\partial T^{(0)}}{\partial k_Q^{\rho}}\right)\frac{(\slashed p_Q-\slashed l+m_Q)}{(p_Q-l)^2-m_Q^2}\gamma ^{\rho} u_Q\right| _{k_q=k',k_Q=K'}\\
&k'=p_{\bar{q}}+\alpha l,\;\;K'=p_Q-\alpha l\\
\end{split}
\label{eq.3.13}
\end{equation}
We use $\Phi _{\rm Wfc}$ to denote the correction shown in Fig.~\ref{Fig2:epsart}.c. We find
\begin{equation}
\begin{split}
&\Phi_{\rm Wfc}^{(1)}\otimes T^{(0)}=-\frac{g_s^2C_F}{2}\int \frac{d^dl}{(2\pi)^d}\int _0^1d\alpha\int _0^1 d\beta \frac{1}{l^2}\bar{v}_{\bar{q}}\left.\left(\frac{\partial }{\partial k_{q}}-\frac{\partial }{\partial k_Q}\right)^2T^{(0)}\right|_{k_q=k',k_Q=K'}u_Q\\
\end{split}
\label{eq.3.14}
\end{equation}
The corrections shown in Figs.~\ref{Fig2:epsart}d, \ref{Fig2:epsart}e and \ref{Fig2:epsart}f can be denoted as $\Phi _{\rm box}$, $\Phi _{\rm ext Q}$ and $\Phi _{\rm ext q}$. We find that, they have the same forms as the free particle 1-loop QCD corrections.

\section{\label{sec:level5}1-loop factorization}

For simplicity, we denote
\begin{equation}
\begin{split}
&x=m_Q^2,\;\;y=2p_Q\cdot p_{\gamma},\;\;z=2p_{\gamma}\cdot p_{\bar{q}},\;\;w=2p_Q\cdot p_{\bar{q}}
\end{split}
\label{eq.4.1}
\end{equation}
The definitions of $x$, $y$ and $z$ are the same as Ref.~\cite{Genon and Sachrajda Radivative Leptonic} at order $O(\Lambda _{\rm QCD}\left/m_Q\right.)^0$, while $w$ is a new scalar that appears at the order of $O(\Lambda _{\rm QCD}\left/ m_Q\right.)$ contributions which will be shown later, it represents the effect of the transverse momentum.

To calculate the hard-scattering amplitude, we need to calculate all 1-loop corrections of $F_a$, $F_b$ and $F_c$. We take a small mass $m_q$ for the light quark to regulate the collinear IR divergences. The soft IR divergences will not appear explicitly in this factorization procedure. We use $\overline{\rm {MS}}$ \cite{msbar} scheme to regulate the ultraviolet (UV) divergences, in $D=4-\epsilon$ demission, we define $N_{\rm UV}$ as
\begin{equation}
\begin{split}
N_{\rm UV}=\frac{2}{\epsilon} - \gamma _E + \log (4\pi)
\end{split}
\label{eq.4.nuv}
\end{equation}

We take the factorization scale the same as the renormalization scale, so we use the same $\mu$ in $F^{(1)}$ and $\Phi^{(1)} \otimes T^{(0)}$.

\subsection{\label{sec:level51}1-loop correction of $T_a^{(0)}$}

\begin{figure}
\includegraphics[scale=0.9]{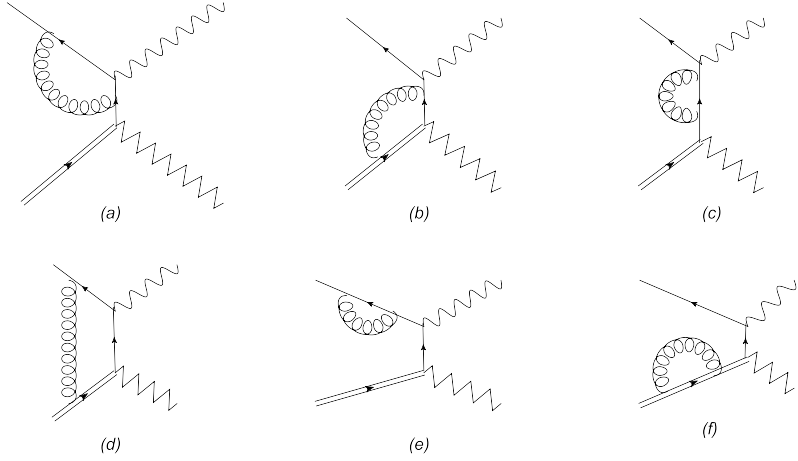}
\caption{\label{Fig3:epsart} 1-loop QCD correction of $F_a$.}
\end{figure}

The Feynman diagrams of the 1-loop corrections of $T_a$ are shown in Fig.~\ref{Fig3:epsart}.

We denote the correction of the electric-magnetic (EM) vertex, the one shown in Fig.~\ref{Fig3:epsart}.a.\, as $F^{(1){\rm EM}}$. To show the effect of the transverse momentum explicitly, we establish the longitudinal part and transverse part separately. The matrix element at tree level can be written as
\begin{equation}
\begin{split}
&F^{(0)}_a=F^{(0)}_{a\parallel}+F^{(0)}_{a\bot},\;\;\;\;F^{(0)}_{a\parallel}=-e_q\bar{v}_{\bar{q}}\frac{\slashed \varepsilon \slashed p_{\gamma}}{2p_{\gamma}\cdot p_{\bar{q}}} P_L^{\mu}u_Q, \;\;\;\;F^{(0)}_{a\bot}= e_q \bar{v}_{\bar{q}} \frac{2\varepsilon \cdot p_{\bar{q}}}{2p_{\gamma}\cdot p_{\bar{q}}} P_L^{\mu}u_Q
\end{split}
\label{eq.4.t0sep}
\end{equation}
The transverse part is at order $O(\left.\Lambda _{\rm QCD}\right/m_Q)$. The corrections to each part is represented separately as
\begin{equation}
\begin{split}
&F^{(1){\rm EM}}_a=F^{(1){\rm EM}}_{a\parallel}+F^{(1){\rm EM}}_{a\bot}\\
&F^{(1){\rm EM}}_{a\parallel}=C_Fg_s^2\int \frac{d^d l}{(2\pi)^d}\bar{v}_{\bar{q}}i\gamma _{\rho}\frac{i(-\slashed p_{\bar{q}}-\slashed l + m_q)}{(p_{\bar{q}}+l)^2-m_q^2}(-ie_q)\slashed \varepsilon \frac{i(-\slashed p_{\bar{q}}-\slashed l + \slashed p_{\gamma}+m_q)}{(p_{\bar{q}}+l-p_{\gamma})^2-m_q^2}i\gamma ^{\rho}\frac{i\slashed p_{\gamma}}{2p_{\bar{q}}\cdot p_{\gamma}}P_L^{\mu}\frac{-i}{l^2}u_Q\\
&F^{(1){\rm EM}}_{a\bot}=ie_qC_Fg_s^2\int \frac{d^d l}{(2\pi)^d}\bar{v}_{\bar{q}}\gamma _{\rho}\frac{(\slashed p_{\bar{q}}+\slashed l - m_q)}{(p_{\bar{q}}+l)^2-m_q^2}\slashed \varepsilon \frac{(\slashed p_{\bar{q}}+\slashed l - \slashed p_{\gamma}-m_q)}{(p_{\bar{q}}+l-p_{\gamma})^2-m_q^2}\gamma ^{\rho}\frac{\slashed p_{\bar{q}}}{2p_{\bar{q}}\cdot p_{\gamma}}P_L^{\mu}\frac{1}{l^2}u_Q\\
\end{split}
\label{eq.4.1.1}
\end{equation}
After performing the momentum-integration, the result is
\begin{equation}
\begin{split}
&F^{(1){\rm EM}}_{a\parallel}=F^{(0)}_{a\parallel} \frac{\alpha _s C_F}{4\pi}\left(N_{{\rm UV}}-\log(\frac{2p_{\gamma}\cdot p_{\bar{q}}}{\mu^2})+2\log(\frac{2p_{\gamma}\cdot p_{\bar{q}}}{m_q^2})\right)\\
&F^{(1){\rm EM}}_{a\bot}=F^{(0)}_{a\bot}\frac{\alpha _s C_F}{4\pi}\left(N_{{\rm UV}} -\log(\frac{2p_{\bar{q}}\cdot p_{\gamma}}{\mu^2}) +1\right)
\end{split}
\label{eq.4.1.2}
\end{equation}
where $C_F$ is defined as $C_F=(N^2-1)\left/2 N\right. = 4\left/3\right.$ for QCD. The transverse part $F_{\bot}^{(1)}$ is an order $O(\alpha _s \Lambda _{\rm QCD}\left/m_Q\right.)$ contribution. Accordingly the 1-loop correction of the wave-function is $\Phi _q^{(1)}\otimes T^{(0)}_a$, which can also be written as $\Phi _q^{(1)}\otimes (T_{a\parallel}^{(0)}+T_{a\bot}^{(0)})$:
\begin{equation}
\begin{split}
&\Phi_q ^{(1)}\otimes T^{(0)}_{a\parallel}=ie_qC_Fg_s^2\int \frac{d^d l}{(2\pi)^d}\frac{1}{l^2}\bar{v}_{\bar{q}}\slashed p_{\gamma}\frac{(\slashed {l}+\slashed {p}_{\bar{q}}-m_q)}{(l+p_{\bar{q}})^2-m_{\bar{q}}^2}\frac{2\slashed p_{\gamma}\slashed \varepsilon_{\gamma}^*}{4p_{\gamma}\cdot (p+l)p_{\gamma}\cdot p}P_L^{\mu } u_Q\\
&\Phi_q ^{(1)}\otimes T^{(0)}_{a\bot}=-ie_qg_s^2C_F\int _0^1 d\alpha\int \frac{d^d l}{(2\pi)^d} \frac{1}{l^2}\\
&\times \bar{v}_{\bar{q}}\frac{4\slashed \varepsilon_{\gamma}(p_{\gamma}\cdot (p_{\bar{q}}+\alpha l)-4\slashed p_{\gamma}\varepsilon _{\gamma}\cdot (p_{\bar{q}}+\alpha l)}{(2p_{\gamma}\cdot (p_{\bar{q}}+\alpha l))^2}\frac{(\slashed {l}+\slashed {p}_{\bar{q}} - m_{\bar{q}})}{(l+p_{\bar{q}})^2-m_{\bar{q}}^2}P_{L\mu}u_Q \\
\end{split}
\label{eq.4.1.3}
\end{equation}
$\Phi _q^{(1)} \otimes T_{\bot}^{(0)}$ is calculated in the light-cone coordinate (see also the Appendix.A in Ref. \cite{TM Yan work}), the result is
\begin{equation}
\begin{split}
&\Phi_q ^{(1)}\otimes T^{(0)}_{a\parallel}=2\frac{C_F\alpha_s}{4\pi}T_{a\parallel}^{(0)}\left(N_{{\rm UV}}-\log \frac{m_q^2}{\mu^2}+2\right),\;\;\Phi_q ^{(1)}\otimes T^{(0)}_{a\bot}=0\\
\end{split}
\label{eq.4.1.4}
\end{equation}
Using Eq.~(\ref{eq.3.2}), we can obtain the hard-scattering kernel
\begin{equation}
\begin{split}
&T_{a}^{(1){\rm EM}}=T^{(0)}_{a\parallel}\frac{C_F\alpha_s}{4\pi}\left(\log \frac{2k_{\bar{q}}\cdot p_{\gamma}}{\mu^2}-4\right)+T^{(0)}_{a\bot}\frac{C_F\alpha_s}{4\pi}\left(-\log \frac{2k_{\bar{q}}\cdot p_{\gamma}}{\mu^2}+1\right)
\end{split}
\label{eq.4.1.6}
\end{equation}
We find that, the hard-scattering kernel of the longitudinal direction is the same as the result in Ref.~\cite{Genon and Sachrajda Radivative Leptonic}.

The correction of the weak vertex is shown in Fig.~\ref{Fig3:epsart}.b. The result can be written as \cite{trangle calc}
\begin{equation}
\begin{split}
&F_a^{(1){\rm weak}}=-ie_qC_Fg_s^2 \frac{i}{16\pi ^2}\bar{v}_{\bar{q}}\slashed \varepsilon \left\{\frac{\slashed p_{\gamma}-\slashed p_{\bar{q}}}{-z}\left[\left(N_{\rm UV}-\log \frac{y}{\mu^2}+y_1-2\log \frac{xz}{y^2}+\frac{x}{x-y}\log\frac{x}{y}\right)\right.\right.\\
&\left.+\frac{w y}{(x-y)^2 y^2}\left(x^2\left(-4\log\frac{xz}{y^2}-2\right)+xy\left(5\log\frac{x}{y}+8\log \frac{z}{y}+5\right)+y^2\left(2\log\frac{y}{x}+4\log\frac{y}{z}-3\right)\right)\right.\\
&\left.+\frac{x^3z}{(x-y)^2 y^2}\left(-3y_1+8\log\frac{xz}{y^2}-4\right)+\frac{x^2yz}{(x-y)^2 y^2}\left(7y_1+13\log\frac{y}{x}+19\log\frac{y}{z}+14\right)\right.\\
&\left.\left.+\frac{xy^2z}{(x-y)^2 y^2}\left(-5y_1+4\log\frac{x}{y}+14\log\frac{z}{y}-17\right)+\frac{y^3z}{(x-y)^2 y^2}\left(y_1-3\log\frac{z}{y}+2\log\frac{x}{y}+7\right)\right]P_L^{\mu}\right.\\
&\left.+\frac{\slashed p_{\gamma}-\slashed p_{\bar{q}}}{-z}4p_{\bar{q}}^{\mu}\slashed p_Q(1-\gamma _5)\frac{\log\frac{x}{y}}{2(x-y)}-2 m_QP_R^{\mu} \frac{1}{y}\log\frac{xz}{y^2}\right\}u_Q\\
\end{split}
\label{eq.4.1.9}
\end{equation}
with
\begin{equation}
\begin{split}
&y_1=-\pi^2+2{\rm Li}_2\left(1-\frac{x}{y}\right)-\log^2\frac{xz}{y^2}+\log^2\frac{x}{y}\\
\end{split}
\label{eq.4.1.y1}
\end{equation}
In the expressions above, we have already neglected the terms which are irrelevant for the discussion, because they will not contribute to the matrix element up to order $O(\left.\Lambda _{\rm QCD}\right/m_Q)$ when convoluted with the distribution amplitudes, due to their Dirac structures \cite{Genon and Sachrajda Radivative Leptonic}. In the rest of this section, we will always use this simplification if possible.

The correction of wave-function is $\Phi_Q ^{(1)}\otimes T_a^{(0)}$, which can be written as
\begin{equation}
\begin{split}
&\Phi^{(1)}_Q\otimes T^{(0)}_{a\parallel}=-2ie_qC_Fg_s^2\bar{v}_{\bar{q}}\slashed \varepsilon \frac{\slashed p_{\gamma}}{2p_{\bar{q}}\cdot p_{\gamma}}P_L^{\mu}\frac{i}{16\pi^2}\left\{\left((\log \frac{z}{y}-1)(1-\frac{z}{y})+\frac{m_Q\slashed p_{\gamma}}{y}\right)N_{\rm UV}\right.\\
&\left.-\left(\left(2-\frac{\pi^2}{3}+(\frac{1}{4}\log^2\frac{x}{\mu^2}-\log^2\frac{y\mu}{zm_Q})\right)(1+\frac{z}{y})+(\frac{z}{y}-1)\log \frac{x}{\mu^2}\right)-\left((\frac{m_Q\slashed p_{\gamma}}{y})(\log \frac{x}{\mu^2}-2)\right)\right\}u_Q
\end{split}
\label{eq.4.1.13}
\end{equation}
and
\begin{equation}
\begin{split}
&\Phi^{(1)}_Q\otimes T^{(0)}_{a\bot}=-2ie_qC_Fg_s^2\frac{i}{16\pi^2}\bar{v}_{\bar{q}}P_L^{\mu} \left\{\left(-\frac{m_Q}{y}\left(\log \frac{z}{y}+1\right)\right)N_{\rm UV}\right.\\
&\left.-\frac{m_Q}{y}\left( (\log \frac{x}{\mu^2}-2)\log \frac{y}{z}-\log\frac{x}{\mu^2}+4\right)\right\}\slashed \varepsilon u_Q\\
\end{split}
\label{eq.4.1.14}
\end{equation}

With Eq.~(\ref{eq.2.3}) and Eq.~(\ref{eq.3.2}), the hard-scattering kernel can be obtained through the following equation
\begin{equation}
\begin{split}
&\Phi ^{(0)}\otimes T_{a}^{(1){\rm weak}}=F_a^{(1){\rm weak}} - (\Phi ^{(1)}\otimes T_a^{(0)})\\
\end{split}
\label{eq.4.1.15}
\end{equation}

The correction in Fig.~\ref{Fig3:epsart}.c.\ is
\begin{equation}
\begin{split}
&F^{(1){\rm wfc}}_a=-\frac{C_F\alpha _s}{4\pi}F_a^{(0)}(N_{{\rm UV}}-\log(\frac{2p_{\gamma}\cdot p_{\bar{q}}}{\mu ^2})+1)
\end{split}
\label{eq.4.1.7}
\end{equation}
We find that the correction of the wave-function also vanishes as Ref.~\cite{Genon and Sachrajda Radivative Leptonic}, i. e.
\begin{equation}
\begin{split}
&\Phi _{\rm Wfc}^{(1)}\otimes T^{(0)}_a=0
\end{split}
\label{eq.4.1.8A}
\end{equation}
Then we obtain
\begin{equation}
\begin{split}
&T^{{\rm wfc}(1)}=\frac{\alpha _s C_F}{4\pi}\left(T_{a\parallel}^{(0)}+T_{a\bot}^{(0)}\right)\left(\log \frac{2p_{\gamma}\cdot k_{\bar{q}}}{\mu ^2} - 1\right)\\
\end{split}
\label{eq.4.1.8}
\end{equation}

The corrections of the external legs and the box diagram are equal to the relevant corrections to the wave-functions because we use the renormalization scale equal to the factorization scale, so
\begin{equation}
\begin{split}
&T^{(1){\rm ext q}}_a=T^{(1){\rm ext Q}}_a=T^{(1){\rm box}}_a=0
\end{split}
\label{eq.4.1.16}
\end{equation}

\subsection{\label{sec:level52}1-loop correction of $T_b^{(0)}$}

\begin{figure}
\includegraphics[scale=0.9]{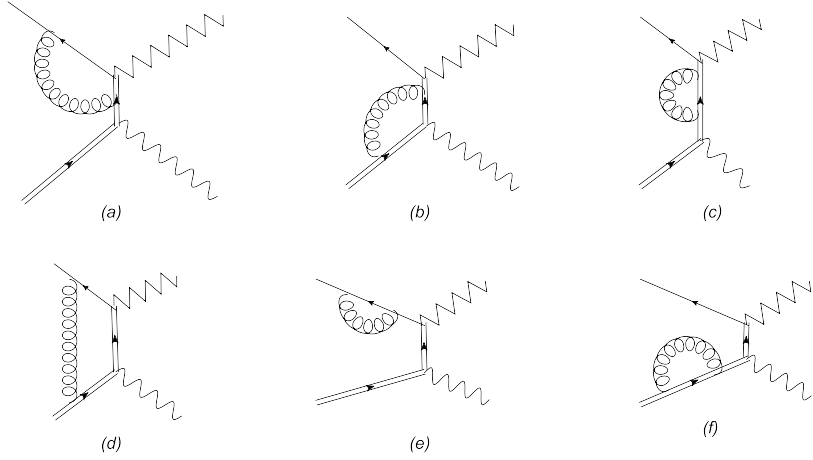}
\caption{\label{Fig4:epsart} 1-loop QCD correction of $F_b$.}
\end{figure}

The corrections of $T_b$ are also order $O(\Lambda _{\rm QCD}\left/m_Q\right.)$ contributions as the tree level, the Feynman diagrams of the 1-loop corrections of $T_b$ are shown in Fig.~\ref{Fig4:epsart}.

The correction of the weak vertex is
\begin{equation}
\begin{split}
&F^{(1){\rm weak}}_b=-ie_QC_Fg_s^2 \frac{i}{16\pi ^2}\bar{v}_{\bar{q}}\left\{\left[N_{\rm UV}-\log\frac{y}{\mu^2}-2+\frac{x}{y-x}+\frac{x^2}{(x-y)^2}\log\frac{x}{y}\right.\right.\\
&\left.\left.+\frac{2wyy_2}{w-z}+\frac{wy\log\frac{x}{y}(3\log\frac{x}{x-y}+\log\frac{y}{x-y})}{(w-z)^2}\right]P_L^{\mu}\frac{\slashed p_{\gamma}}{y}\right.\\
&\left.+2P_L^{\mu}\slashed p_q\frac{2yy_2(w-z)+y\left(\log\frac{y}{x}\left(3\log\frac{x}{x-y}+\log\frac{y}{x-y}\right)\right)}{2 (w-z)^2}\right\}\slashed \varepsilon u_Q\\
\end{split}
\label{eq.4.2.2}
\end{equation}
with
\begin{equation}
\begin{split}
&y_2=\frac{1}{2 (w-z)}\left( \log\left(\left(\frac{x}{w+x-y-z}\right)^3\frac{y-w+z}{x-y+w-z}\right)\log\left(\frac{x}{y-w+z}\right)\right)\\
\end{split}
\label{eq.4.2.y2}
\end{equation}
and the correction of the wave-function is
\begin{equation}
\begin{split}
&\Phi _q^{(1)}\otimes T_{b}^{(0)}\sim O(\frac{1}{m_Q^2})\\
\end{split}
\label{eq.4.2.1orderwavefunctionweak}
\end{equation}
the hard kernel can be obtained using
\begin{equation}
\begin{split}
&\Phi ^{(0)}\otimes T_b^{{\rm weak}(1)}=F_b^{{\rm weak}(1)}\\
\end{split}
\label{eq.4.2.3}
\end{equation}
The results of the EM vertex is
\begin{equation}
\begin{split}
&F^{(1){\rm EM}}_b=-ie_QC_Fg_s^2\bar{v}_{\bar{q}}\frac{i}{16\pi^2}P_L^{\mu}\left\{\frac{m_Q\slashed \varepsilon }{y}\frac{-2y}{x-y}\log \frac{x}{y}+\frac{\slashed p_{\gamma}\slashed \varepsilon }{y}\left(2N_{\rm UV}-\log \frac{x}{\mu^2} \right.\right.\\
&\left.\left.-\frac{x}{y}\left(-2{\rm Li}_2(1-\frac{y}{x})+\frac{\pi^2}{3}\right)  - \frac{6x-y}{x-y}\log \frac{x}{y} \right)\right\}u_Q\\
&\Phi _Q^{(1)}\otimes T_b^{(0)}=-2ie_QC_Fg_s^2 \frac{i}{16\pi^2}\bar{v}_{\bar{q}}P_L^{\mu} \frac{\slashed p_{\gamma}\slashed \varepsilon}{y}u_Q\left(N_{\rm UV}-\log\frac{x}{\mu^2}+2\right)\\
&T_b^{(1){\rm EM}}=T_b^{(0)}\frac{C_Fg_s^2}{4\pi}\left(\log \frac{m_Q^2}{\mu^2}-4-\frac{m_Q^2}{2p_{\gamma}\cdot k_Q}\left(-2{\rm Li}_2(1-\frac{2p_{\gamma}\cdot k_Q}{m_Q^2} )+\frac{\pi^2}{3}\right)\right.\\
&\left.-\frac{6m_Q^2-2p_{\gamma}\cdot k_Q}{m_Q^2-2p_{\gamma}\cdot k_Q}\log \frac{m_Q^2}{2p_{\gamma}\cdot k_Q}\right)\\
&+ie_QC_Fg_s^2\frac{i}{16\pi^2}\bar{v}_{\bar{q}}\frac{P_L^{\mu}\slashed \varepsilon m_Q}{2p_{\gamma}\cdot k_Q}u_Q\times \frac{4p_{\gamma}\cdot k_Q}{m_Q^2-2p_{\gamma}\cdot k_Q}\log \frac{m_Q^2}{2p_{\gamma}\cdot k_Q}\\
\end{split}
\label{eq.4.2.1}
\end{equation}
The results in Fig.~\ref{Fig4:epsart}.c.\ is
\begin{equation}
\begin{split}
&F_{b}^{(1){\rm wfc}}=-ie_QC_fg_s^2\frac{i}{16\pi^2}\bar{v}_{\bar{q}}P_L^{\mu}\left\{\left(-N_{UV}-1+\log\frac{x}{\mu^2}-\frac{x}{(x-y)}+
\frac{y(2x-y)}{(x-y)^2}\log\frac{x}{y}\right.\right.\\
&\left.\left.+\frac{2x}{y}(-3N_{UV}-5+3\log\frac{x}{\mu^2}+\frac{x}{x-y}+\frac{(2x-3y)y}{(x-y)^2}\log\frac{x}{y}) \right) \frac{\slashed p_{\gamma}}{y}\slashed \varepsilon u_Q\right. \\
&\left.+\left(-3N_{UV}-5+3\log\frac{x}{\mu^2}+\frac{x}{x-y}+\frac{(2x-3y)y}{(x-y)^2}\log\frac{x}{y}\right)\frac{m_Q\slashed \varepsilon }{y} u_Q\right\}\\
&\Phi ^{(1)}_{\rm Wfc}\otimes T_b^{(0)}=0\\
&T_b^{(1){\rm wfc}}=\frac{\alpha _s C_F}{4\pi}T_{b\parallel}^{(0)}\left(-1+\log\frac{x}{\mu^2}-\frac{x}{x-y}+\frac{y(2x-y)}{(x-y)^2}\log\frac{x}{y}\right.\\
&\left.+\frac{2x}{y}(-5+3\log\frac{x}{\mu^2}+\frac{x}{x-y}+\frac{(2x-3y)y}{(x-y)^2}\log\frac{x}{y}) \right) \\
&-ie_QC_Fg_s^2\frac{i}{16\pi^2}\bar{v}_{\bar{q}}P_L^{\mu}\slashed \varepsilon u_Q\left(\frac{m_Q}{y}(-5+3\log\frac{x}{\mu^2}+\frac{x}{x-y}+\frac{y(2x-3y)}{(x-y)^2}\log\frac{x}{y})\right)\left.\right|_{p_Q\to k_Q}
\end{split}
\label{eq.4.2.4}
\end{equation}
And the correction of the external legs and the box correction are also equal to each other, so we also have
\begin{equation}
\begin{split}
&T^{(1){\rm ext q}}_b=T^{(1){\rm ext Q}}_b=T^{(1){\rm box}}_b=0
\end{split}
\label{eq.4.2.5}
\end{equation}

\subsection{\label{sec:level53}1-loop correction of $T_c^{(0)}$}

The correction of the distribution function with gluons in Wilson-Line does not have correspondent 1-loop QCD corrections. And because the momentum of light-quark and heavy-quark show up together as $k_Q+k_{\bar{q}}$ in $T_c^{(0)}$, we find
\begin{equation}
\begin{split}
&\left(\frac{\partial }{\partial k_Q}-\frac{\partial }{\partial k_{\bar{q}}}\right)T_c^{(0)}=0
\end{split}
\label{eq.4.3.1}
\end{equation}
and we obtain
\begin{equation}
\begin{split}
&T_c^{(1)q}=T_c^{(1)Q}=T_c^{(1){\rm wfc}}=0\\
\end{split}
\label{eq.4.3.2}
\end{equation}

All the other corrections are similar as Eq.~(\ref{eq.4.1.16}) and Eq.~(\ref{eq.4.2.5}), we find
\begin{equation}
T_c^{(1){\rm extq}}=T_c^{(1){\rm extQ}}=T_c^{(1){\rm trangle}}=0
\label{eq.4.3.3}
\end{equation}

\subsection{\label{sec:level54}1-loop result summery}

With Eqs.~(\ref{eq.4.1.6}), (\ref{eq.4.1.15}), (\ref{eq.4.1.8}), (\ref{eq.4.1.16}), (\ref{eq.4.2.3}) - (\ref{eq.4.2.5}), (\ref{eq.4.3.2}) and (\ref{eq.4.3.3}), we can establish the order $\alpha _s$ hard-scattering kernel. We find that, $T^{(1)}$ is IR finite up to order $O(\alpha _s \Lambda _{\rm QCD}\left/ m_Q\right.)$, so the factorization is proved up to the order of $O(1\left/m_Q\right.)$ corrections.

It is well known that, the factorization will fail with the light mesons. However we find that the $\Lambda _{\rm QCD}\left.\right/m_Q$ expansion is irrelevant by briefly investigating the factorization at higher orders of $O(\Lambda _{\rm QCD}\left.\right/m_Q)$.

In the calculations above, we find that the IR divergences in the corrections of the external legs and the box diagrams are cancelled exactly, while the remaining IR divergences are all collinear divergences which show up in the amplitudes with one of the vertexes of the gluon propagator connecting to the external light-quark, which are denoted in Fig~\ref{Fig3:epsart}.a and Fig~\ref{Fig4:epsart}.a. In $T_a^{(1){\rm EM}}$, there are no neglected $O(\Lambda _{\rm QCD}\left/m_Q\right.)^2$ or higher order contributions. As a result, the IR divergences at higher orders only survive in $T_b^{(1){\rm weak}}$. We also find that, there are both higher order IR divergences in $F_b^{(1){\rm weak}}$ and $\Phi _q^{(1)} \otimes T_b^{(0)}$. We shall investigate whether those IR divergences can be cancelled.

We concentrate on the IR region of the loop integration, using $l_{\epsilon}\to 0$, we find
\begin{equation}
\begin{split}
&F_{\rm b\;IR}^{\rm weak(1)}=\lim _{\substack{l_{\epsilon}\to 0}}\left(-ie_QC_Fg_s\int _{-l_{\epsilon}}^{l_{\epsilon}} \frac{d^d l}{(2\pi)^d} \bar{v}_q \frac{1}{l^2} \gamma _{\rho} \frac{-\slashed p_{\bar{q}}}{(l+p_{\bar{q}})^2-m_q^2}P_L^{\mu}\right.\\
&\left.\times \frac{2p_Q^{\rho}\slashed p_Q+2p_{\gamma}^{\rho}\slashed p_{\gamma}-2 p_Q^{\rho} \slashed p_{\gamma}-2 p_{\gamma}^{\rho} \slashed p_Q+2 \gamma^{\rho} p_Q\cdot p_{\gamma}+2(p_Q-p_{\gamma})^{\rho}m_Q}{(2p_Q\cdot p_{\gamma})^2} \slashed \varepsilon u_Q\right)\\
\end{split}
\label{eq.6.1}
\end{equation}
We also bring back the neglected higher order terms in the numerator of $T_b^{(0)}$, and the correction of the wave-function is
\begin{equation}
\begin{split}
&T_b^{(0)}=e_QP_L^{\mu}\frac{\slashed k_Q -\slashed p_{\gamma}+m_Q}{(k_Q-p_{\gamma})^2-m_Q^2}\slashed \varepsilon\\
&(\Phi _q^{(1)} \otimes T^{(0)})_{IR}=\lim _{\substack {l_{\epsilon}\to 0}}\left(-ie_QC_Fg_s^2 \int_{-l_{\epsilon}}^{l_{\epsilon}} \frac{d^dl }{(2\pi)^d} \frac{1}{l^2} \bar{v}_q \gamma ^{\rho} \frac{\slashed p_{\bar{q}}}{(l+p_{\bar{q}})^2-m_q^2}P_L^{\mu}\right.\\
&\left.\times \frac{-\gamma _{\rho}2p_Q\cdot p_{\gamma} -2(p_Q-p_{\gamma})^{\rho}(\slashed p_Q-\slashed p_{\gamma}+m_Q)}{(2p_Q\cdot p_{\gamma})^2}\slashed \varepsilon u_Q\right)
\end{split}
\label{eq.6.2}
\end{equation}
we find
\begin{equation}
\begin{split}
&F^{(1)}_{\rm b\;IR}-(\Phi ^{(1)}\otimes T^{(0)})_{IR}=0\\
\end{split}
\label{eq.6.3}
\end{equation}
This result indicates that, the factorization is valid at any order of $O(\Lambda _{\rm QCD}\left/m_Q\right.)$, as a result, the valid region of factorization in Ref \cite{Genon and Sachrajda Radivative Leptonic} is extended. The failure of the factorization approach in the light meson decays is due to the bad convergence behaviour in the $O(\alpha _s)$ expansion.

We then concentrate on the contribution of the hard scattering kernel to the amplitude. The amplitude can be obtained by replacing the wave-function with the one obtained in Ref.~\cite{Teacher Yang wave function}
\begin{equation}
\begin{split}
&\Phi ^{(0)}(k_q, k_Q)=\frac{1}{\sqrt{3}}\int d^3 k \Psi (k) \frac{1}{\sqrt{2}}M\mid 0> \\
&\times \delta ^3(\vec{k}_{\bar{q}} + \vec{k})\delta ^3(\vec{k}_Q - \vec{k})\delta (k_{\bar{q}0}-\sqrt{k^2+m_q^2})\delta (k_{Q0}-\sqrt{k^2+m_Q^2})
\end{split}
\label{eq.7.1}
\end{equation}
with
\begin{equation}
\begin{split}
&M=\sum _{\substack{i}}b_Q^{i+}(\vec{k},\uparrow)d_q^{i+}(-\vec{k},\downarrow)-b_Q^{i+}(\vec{k},\downarrow)d_q^{i+}(-\vec{k},\uparrow), \;\;\;\;\Psi (\vec{k})=4\pi \sqrt{m_P \lambda _P ^3} e^{-\lambda _P |\vec{k}|}
\end{split}
\label{eq.7.2}
\end{equation}
The matrix element can be written as
\begin{equation}
\begin{split}
&F^{\mu}(\mu)=\frac{1}{(2\pi)^3}\frac{3}{\sqrt{6}}\int d^3k \int d^4 k_Q \int d^4 k_{\bar{q}} \Psi (k)  Tr\left[M\cdot \left(T^{(0)\mu}(k_{\bar{q}}, k_Q)+T^{(1)\mu}(k_{\bar{q}}, k_Q)\right)\right]\\
&\times \delta ^3(\vec{k}_{\bar{q}} + \vec{k})\delta ^3(\vec{k}_Q - \vec{k})\delta (k_{\bar{q}0}-\sqrt{k^2+m_q^2})\delta (k_{Q0}-\sqrt{k^2+m_Q^2})
\end{split}
\label{eq.7.3}
\end{equation}

After convolution with the wave-functions of the heavy mesons, some of the terms will have identical contributions to the matrix element up to order $O(\Lambda _{\rm QCD}\left/m_Q\right.)$ due to their Dirac structures. As a result, we find that the $F^{\mu}$ can be simplified as a function of four different types of Dirac structures, and can be written as
\begin{equation}
\begin{split}
&F^{\mu}(\mu)=%\sum _{\substack{i}}\frac{1}{(2\pi)^3}\frac{3}{\sqrt{6}}\int d^3k \int d^4 k_Q \int d^4 k_{\bar{q}} \Psi (k) Tr\left[C_i(k_Q, k_{\bar{q}}, \mu) M\cdot K_i(k_Q, k_{\bar{q}})\right]\\
%&\times \delta ^3(\vec{k}_{\bar{q}} + \vec{k})\delta ^3(\vec{k}_Q - \vec{k})\delta (k_{\bar{q}0}-\sqrt{k^2+m_q^2})\delta (k_{Q0}-\sqrt{k^2+m_Q^2})\\
%&=
\sum _{\substack{n}}\frac{1}{(2\pi)^3}\frac{3}{\sqrt{6}}\int d^3k \Psi (k) Tr\left[C_n(p_Q, p_{\bar{q}}, \mu) M\cdot K_n(p_Q, p_{\bar{q}})\right]
\end{split}
\label{eq.4.4.1}
\end{equation}
with $p_q=(\sqrt{m_q^2 + k^2}, -\vec{k})$ and $p_Q=(\sqrt{m_Q^2 +k^2}, \vec{k})$ denote the on-shell momenta of the light anti-quark and the heavy quark in the bound state. And the $K_n$ are defined as
\begin{equation}
\begin{split}
&K_1(k_Q, k_{\bar{q}})=T_{a\parallel}^{(0)},\;\;K_2(k_Q, k_{\bar{q}})=T_{a\bot}^{(0)},\;\;K_3(k_Q, k_{\bar{q}})=T_b^{(0)},\;\;K_4(k_Q, k_{\bar{q}})=e_Q\frac{P_L^{\mu}\slashed \varepsilon m_Q}{2p_{\gamma}\cdot k_Q}\\
\end{split}
\label{eq.4.4.2}
\end{equation}
Except for $C_1K_1$, all the other products are contribution of order $O(\Lambda_{\rm QCD}\left/m_Q\right.)$, for clarity, we define $C_1=C_1^0+C_1^1$, with $C_1^m$ represents order $O(\Lambda _{\rm QCD}\left/m_Q\right.)^m$ contribution, the coefficients are
\begin{equation}
\begin{split}
&C_1^0(p_q,p_Q,\mu)=1+\frac{\alpha _s C_F}{4\pi}\left(-\log \frac{y}{\mu^2}+y_1-2\log \frac{xz}{y^2}+\frac{x}{x-y}\log\frac{x}{y}-4+\frac{2\pi^2}{3}\right.\\
&\left.\left.+2\log ^2\frac{y}{z} -2\log \frac{y}{z} \log \frac{x}{\mu^2} +2\log\frac{x}{\mu^2}+2\log \frac{z}{\mu ^2} - 5\right)\right.\\
\end{split}
\label{eq.4.4.c10}
\end{equation}
\begin{equation}
\begin{split}
&C_1^1(p_q,p_Q,\mu)=\frac{w }{(x-y)^2 y}\left(x^2\left(-4\log\frac{xz}{y^2}-2\right)+xy\left(5\log\frac{x}{y}+8\log \frac{z}{y}+5\right)\right.\\
&\left.+y^2\left(2\log\frac{y}{x}+4\log\frac{y}{z}-3\right)\right)\\
&+\frac{x^3z}{(x-y)^2 y^2}\left(-3y_1+8\log\frac{xz}{y^2}-4\right)+\frac{x^2z}{(x-y)^2 y}\left(7y_1+13\log\frac{y}{x}+19\log\frac{y}{z}+14\right)\\
&\left.+\frac{xz}{(x-y)^2}\left(-5y_1+4\log\frac{x}{y}+14\log\frac{z}{y}-17\right)+\frac{yz}{(x-y)^2}\left(y_1-3\log\frac{z}{y}+2\log\frac{x}{y}+7\right)\right.\\
&\left.-\frac{w}{2(x-y)}\log\frac{x}{y}+\left(\frac{2w}{y}-\frac{4xz}{y^2}\right)\log\frac{xz}{y^2}+\left(\frac{2w}{y}-\frac{4xz}{y^2}\right)\left(  \log \frac{x}{\mu^2}\log \frac{y}{z}-2\log \frac{y}{z}-\log\frac{x}{\mu^2}+4\right)\right.\\
&-\left(2-\frac{\pi^2}{3}-\log ^2\frac{y}{z} +\log \frac{y}{z} \log \frac{x}{\mu^2}+\log \frac{x}{\mu^2} \right)\frac{2z}{y}\\
\end{split}
\label{eq.4.4.c11}
\end{equation}
\begin{equation}
\begin{split}
&C_2(p_q,p_Q,\mu)=1+\frac{\alpha _s C_F}{4\pi}\left(-\log \frac{y}{\mu^2}+y_1-2\log \frac{xz}{y^2}+\log\frac{x}{y}+\frac{xz}{yw}\log\frac{xz}{y^2}\right.\\
&\left.-\frac{4zx}{yw}\left( \log \frac{x}{\mu^2}\log \frac{y}{z}-2\log \frac{y}{z}-\log\frac{x}{\mu^2}+4\right)\right)\\
\end{split}
\label{eq.4.4.c2}
\end{equation}
\begin{equation}
\begin{split}
&C_3(p_q,p_Q,\mu)=1+\frac{\alpha _s C_F}{4\pi}\left(\left(\log \frac{x}{\mu^2}-2-\frac{2x}{y}\left(-{\rm Li}_2(1-\frac{y}{x})+\frac{\pi^2}{3}\right)-\frac{6x-y}{x-y}\log \frac{x}{y}\right)\right.\\
&\left.-1+\log\frac{x}{\mu^2}-\frac{x}{x-y}+\frac{y(2x-y)}{(x-y)^2}\log\frac{x}{y}+\frac{2x}{y}(-5+3\log\frac{x}{\mu^2}+\frac{x}{x-y}+\frac{(2x-3y)y}{(x-y)^2}\log\frac{x}{y}) \right.\\
&\left.-\log\frac{y}{\mu^2}-2+\frac{x}{y-x}+\frac{x^2}{(x-y)^2}\log\frac{x}{y}+\frac{2wyy_2}{w-z}+\frac{wy\log\frac{x}{y}(3\log\frac{x}{x-y}+\log\frac{y}{x-y})}{(w-z)^2}\right)\\
\end{split}
\label{eq.4.4.c3}
\end{equation}
\begin{equation}
\begin{split}
&C_4(p_q,p_Q,\mu)=\frac{\alpha _s C_F}{4\pi}\left(\left(\frac{2xz^2}{yw}-2z\right)\frac{2 y_2 y (w-z)+\log \frac{y}{x} (3 y \log \frac{x}{x-y}+y \log \frac{y}{x-y})}{(w-z)^2}\right.\\
&\left.-\frac{2y}{x-y}\log\frac{x}{y}+\left(-5+3\log\frac{x}{\mu^2}+\frac{x}{x-y}+\frac{y(2x-3y)}{(x-y)^2}\log\frac{x}{y}\right)\right)\\
\end{split}
\label{eq.4.4.c4}
\end{equation}
with $x$, $y$, $z$ and $w$ defined in Eq.~(\ref{eq.4.1}), and $y_1$, $y_2$ defined in Eq.~(\ref{eq.4.1.y1}) and Eq.~(\ref{eq.4.2.y2}).

\section{\label{sec:level6}Large logarithm resummation}

In the expression of $F^{\mu}$, large logarithms show up. Those large logarithms need to be resummed so that the result is phenomenologically reliable. We concentrate on the large logarithms at order $O(\Lambda _{\rm QCD}\left/m_Q\right.)^0$, because terms like $\left(\Lambda _{\rm QCD}\log \frac{\sqrt{m_Q\Lambda _{QCD}}}{m_Q}\right)\left/m_Q\right.$ are suppressed and not large. In this section, we use $y=2p_{\gamma}\cdot p_Q = 2E_{\gamma} m_Q$ for simplicity. We concentrate on $C_1^0$, which can also be written as
\begin{equation}
\begin{split}
&C_1^0(p_q,p_Q,\mu)=1+\frac{\alpha _s(\mu)C_F}{4\pi}\left(-2{\rm Li}_2\left(1-\frac{y}{x}\right)-2\log^2\frac{x}{y}-\frac{y}{x-y}\log\frac{y}{x}+2\log\frac{y}{x}\right.\\
&+2\log\frac{x}{y}\log\frac{x}{\mu^2}+3\log\frac{x}{\mu^2}-\log^2\frac{x}{\mu^2}- 6-\frac{\pi^2}{12}\left.+\log^2\frac{\mu^2}{z}- 3-\frac{\pi^2}{4}\right)\\
\end{split}
\label{eq.5.1}
\end{equation}
When $\mu \sim m_Q$, the large logarithms are the same as Ref.~\cite{Genon and Sachrajda Radivative Leptonic}. However, when $\mu$ is set to $\mu \sim \sqrt{m_Q\Lambda _{QCD}}$, the large logarithms are different from Ref.~\cite{Genon and Sachrajda Radivative Leptonic} by
\begin{equation}
\begin{split}
&-\frac{1}{2}\log^2\frac{m_Q}{\mu^2}+\frac{1}{2}\log \frac{m_Q}{\mu^2}
\end{split}
\label{eq.5.2}
\end{equation}
This is because the corrections of wave-functions in this work are not calculated in HQET framework as Ref.~\cite{Genon and Sachrajda Radivative Leptonic}, in which $m_Q$ will not appear in the result.

\subsection{\label{sec:level62}Renormalization group equation (RGE) evolution}

We start with the RGE at order $O(\Lambda _{\rm QCD}\left/m_Q\right.)^0$, and use the method introduced in Ref.~\cite{SCET work 1}. The RGE of $C_1^0$ is
\begin{equation}
\begin{split}
&\mu \frac{\partial }{\partial \mu}C_1^0(\mu)=\gamma (\mu)C_1^0(\mu)\\
\end{split}
\label{eq.5.6}
\end{equation}
with $\gamma (\mu)$ defined as anomalous dimension of $C_1^0$, which can be obtained by using the counter terms of $C_1^0K_1$. The counter terms at the order $O(\Lambda _{\rm QCD}\left/m_Q\right.)^0$ can be written as
\begin{equation}
\begin{split}
&Z=\frac{\alpha _s(\mu) C_F}{4\pi}\frac{2}{\epsilon}\left(2\log \frac{z}{y}-3\right)
\end{split}
\label{eq.5.3}
\end{equation}
We can use the counter terms and the $\beta$ function in QCD to calculate $\gamma (\mu)$
\begin{equation}
\begin{split}
&\beta = -g \frac{\epsilon}{2}+O(g^3)\;\;\;\;\;\gamma (\mu) = Z ^{-1}\left(\mu \frac{\partial }{\partial \mu} + \beta \frac{\partial}{\partial g}\right)Z \\
\end{split}
\label{eq.5.4}
\end{equation}
The result is
\begin{equation}
\begin{split}
&\gamma (\mu)= -\frac{\alpha _s(\mu) C_F}{2\pi}\left(2\log \frac{\mu^2}{y}+2\log \frac{z}{\mu^2}-3\right)\\
\end{split}
\label{eq.5.5}
\end{equation}
With this result, we can solve the RGE of the coefficient. Assume the coefficient can be written as a hard function multiplied by a jet function \cite{Neubert work,jet function in resume}, so
\begin{equation}
\begin{split}
&C(\mu)=H(\mu)J(\mu),\;\;\;\;\; \gamma (\mu) = \gamma _H (\mu)+\gamma _J (\mu)\\
&\left(\mu \frac{\partial }{\partial \mu}H (\mu)\right)J(\mu) + H (\mu) \left(\mu \frac{\partial }{\partial \mu} J(\mu)\right)=\gamma_H (\mu)H (\mu)J(\mu)+\gamma_J (\mu) H (\mu)J(\mu)\\
\end{split}
\label{eq.5.7}
\end{equation}
The RGE of the hard function is
\begin{equation}
\begin{split}
&\mu \frac{\partial }{\partial \mu}H (\mu)=\gamma _H (\mu) H (\mu)\\
\end{split}
\label{eq.5.8}
\end{equation}
By splitting the coefficient $C^{(0)}_1$ into $H$ and $J$, we assume that the natural scale of $H$ is $m_Q$, while the natural scale of $J$ is $\sqrt{m_Q\Lambda _{QCD}}$, which is also the case for $\gamma (\mu)$ with a $\log (\mu^2 \left/ y\right.)$ term and a $\log (\mu^2 \left/ z\right.)$ term. So we split the $\gamma (\mu)$ such that $\gamma _H(\mu)$ is the sum of the $\log (\mu^2 \left/ y\right.)$ term and an undetermined constant $n'$, so
\begin{equation}
\begin{split}
&\gamma _H(\mu)=-\frac{\alpha _s(\mu)C_F}{2\pi}\left(2\log \frac{\mu^2}{y} + n'\right)\\
\end{split}
\label{eq.5.9}
\end{equation}
Similar as Refs.~\cite{SCET work 2,Genon and Sachrajda Radivative Leptonic}, we find
\begin{equation}
\begin{split}
&\gamma _{H_{\rm LO}} = -\frac{\alpha _s(\mu) C_F}{\pi}2\log \frac{\mu}{m_Q},\;\;\gamma _{H_{\rm NLO}} = -\frac{\alpha _s(\mu)C_F}{2\pi} \left(n'- 2\log\frac{2E_{\gamma}}{m_Q} \right)-2C_FB\frac{\alpha _s^2(\mu)}{(2\pi)^2}\log \frac{\mu}{m_Q}\\
\end{split}
\label{eq.5.10}
\end{equation}
The $\gamma _{H_{\rm NLO}}$ is the same as Refs.~\cite{SCET work 2,Genon and Sachrajda Radivative Leptonic}, so the solution is
\begin{equation}
\begin{split}
&H(\mu)=\exp \left(\frac{f_0}{\alpha _s(m_Q)}+f_1\right) H(m_Q)\\
&f_0=\alpha _s(m_Q) \left(-2\frac{4\pi C_F}{\beta _0^2 \alpha _s(m_Q)}\left(\frac{1}{r}-1+\log r\right)\right)\\
&f_1=-\frac{C_F \beta _1}{\beta _0^3}(1-r+r\log r -\frac{1}{2}\log ^2 r)+\frac{C_F}{\beta _0}\left(n' - 2\log \frac{y}{x}\right)\log r - \frac{2C_F B}{\beta _0^2}(r-1-\log r)
\end{split}
\label{eq.5.11}
\end{equation}
with $r=\frac{\alpha _s (\mu)}{\alpha _s (m_Q)}$, $\beta _0 = \frac{11C_A}{3}-\frac{2N_f}{3}$ and $\beta _1=\frac{34C_A^2}{3}-\frac{10C_AN_f}{3}-2C_FN_f$, where $C_A=3$ for QCD, $N_f$ is the number of the flavour of quarks taken into account, and $B$ can only been derived from 2-loop calculations. In Ref.~\cite{SCET work 2}, by comparing the result with $B\to X_s\gamma$ and $B\to X_ul\bar{v}$ in Ref. \cite{compare B term}, $B$ is found to be $B=C_A\left(\frac{67}{18}-\frac{\pi^2}{6}\right)-\frac{5N_f}{9}$. So the result of $H$ is
\begin{equation}
\begin{split}
H(\mu)=H(m_Q)\exp \left(\frac{\alpha_s(m_Q)C_f}{4\pi}\left(-4\log^2\frac{\mu}{m_Q}+4\log \frac{y}{x}\log \frac{\mu}{m_Q}-2n' \log \frac{\mu}{m_Q}\right) +O(\alpha_s^2)\right)
\end{split}
\label{eq.5.12}
\end{equation}

\subsection{\label{sec:level63}The resummation}

The hard function can be derived by using the method introduced in Ref.~\cite{Neubert work}. With $x_{\gamma}$ defined as $x_{\gamma}=2E_{\gamma}\left/m_Q\right.$, the result is
\begin{equation}
\begin{split}
&\hat {H}(\frac{2E_{\gamma}}{\mu})=1+\frac{\alpha _s(\mu)C_F}{4\pi}\left(-2{\rm Li}_2\left(1-x_{\gamma}\right)-2\log^2\frac{2E_{\gamma}}{\mu}-\frac{1}{1-x_{\gamma}}\log x_{\gamma}+2\log\frac{2E_{\gamma}}{\mu}\right)\\
\end{split}
\end{equation}
and
\begin{equation}
\begin{split}
H(m_Q)=\hat {H}(\frac{2E_{\gamma}}{m_Q})
\end{split}
\end{equation}
we also find
\begin{equation}
\begin{split}
H(m_Q)=C_{3,6}^{SCET}(m_Q)
\end{split}
\end{equation}

With $n' = 3$, the evaluation of the RGE will correctly resum the large logarithms, which can be shown explicitly by expanding the solution of RGE of hard function
\begin{equation}
\begin{split}
&H(\mu)=H(m_Q)\times \left(1+\frac{\alpha _s(m_Q) C_F}{4\pi}\left(-4\log^2\frac{\mu}{m_Q}+4\log\frac{y}{x}\log\frac{\mu}{m_Q}-6\log\frac{\mu}{m_Q}\right)\right)+O(\alpha_s^2)\\
&=1+\frac{\alpha _s(m_Q)C_F}{4\pi}\left(-2{\rm Li}_2\left(1-\frac{y}{x}\right)-2\log^2\frac{x}{y}-\frac{y}{x-y}\log\frac{y}{x}+2\log\frac{y}{x}\right.\\
&\left.+2\log\frac{x}{y}\log\frac{x}{\mu^2}+3\log\frac{x}{\mu^2}-\log^2\frac{x}{\mu^2}- 6-\frac{\pi^2}{12}\right)+O(\alpha_s^2)\\
\end{split}
\label{eq.6.hmq}
\end{equation}
Comparing Eq.~(\ref{eq.5.6}) and Eq.~(\ref{eq.5.7}) with Eq.~(\ref{eq.6.hmq}), we find that there are no more large logarithms in the remaining terms, which gives the jet function
\begin{equation}
\begin{split}
&J(\mu)=1+\frac{\alpha _s(\mu)C_F}{4\pi}\left(\log^2\frac{\mu^2}{z}- 3-\frac{\pi^2}{4}\right)\\
\end{split}
\end{equation}
The $\gamma _J(\mu)$ can been obtained by subtracting $\gamma _H(\mu)$ from $\gamma (\mu)$
\begin{equation}
\begin{split}
&\gamma _J(\mu)=\frac{\alpha _s(\mu) C_F}{\pi} 2\log \frac{\mu }{\sqrt{z}}\\
\end{split}
\end{equation}
The solution of RGE of jet function is
\begin{equation}
\begin{split}
&J(\mu)=J(\sqrt{z})\times \exp \left(\frac{\alpha _s(\sqrt{z})C_F}{4\pi}\log ^2 \frac{\mu^2}{z^2}+O(\alpha _s^2)\right)
\end{split}
\label{eq.7.jetsoluton}
\end{equation}
We find that, Eq.~(\ref{eq.7.jetsoluton}) can also correctly resum the large logarithms which will show up in jet function when $\mu$ is evaluated to lower than $\sqrt{\Lambda _{\rm QCD}m_Q}$.

\begin{figure}
\includegraphics[scale=0.6]{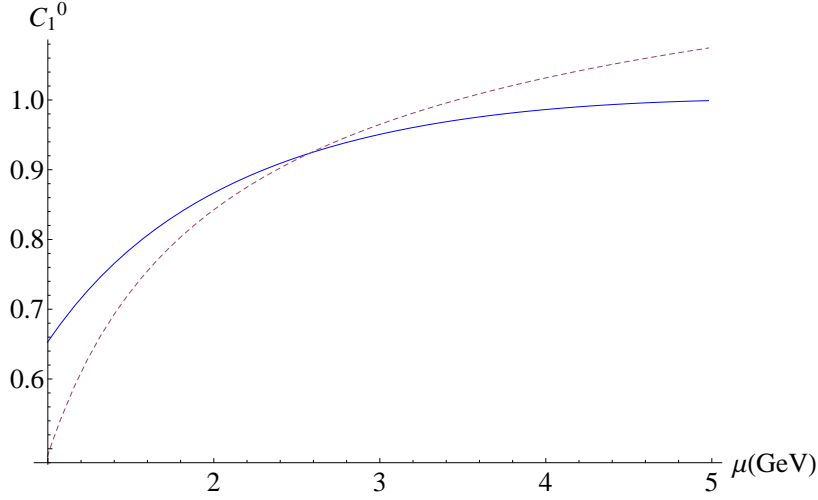}
\caption{\label{Fig5:epsart} $C_1^0$ as function of $\mu$, with $E_{\gamma}=\frac{m_Q}{4}$, $k_+= \Lambda _{\rm QCD}$. The solid line is the resummed result, and the dotted line is the un-resummed one.}
\end{figure}

When $\mu > \sqrt{\Lambda _{\rm QCD} m_Q}$, the resummed result at order $O(\alpha _s \left(\Lambda _{\rm QCD}\left/ m_Q\right.\right)^0)$ is
\begin{equation}
\begin{split}
&C_1^0=\left\{1+\frac{\alpha _s(m_Q)C_F}{4\pi}\left(-2{\rm Li}_2\left(1-\frac{y}{x}\right)-2\log^2\frac{y}{x}-\frac{y}{x-y}\log\frac{y}{x}+2\log\frac{y}{x}-6-\frac{\pi^2}{12}\right)\right\}\\
&\times \exp \left(\frac{\alpha _s(m_Q) C_f}{4\pi}\left(-4\log^2\frac{\mu}{m_Q}+4\log\frac{y}{x}\log\frac{\mu}{m_Q}-6\log\frac{\mu}{m_Q}\right)\right)\\
&\times \left(1+\frac{\alpha _s(\mu) C_f}{4\pi}\left( \log^2 \frac{\mu ^2}{z}- 3-\frac{\pi^2}{4}\right)\right)
\end{split}
\end{equation}
At the leading order of $O(\Lambda _{\rm QCD}\left/m_Q\right.)$, $z$ can be written as $z=2p_{\gamma}\cdot k_{\bar{q}}=\sqrt{2}k_+E_{\gamma}$. Using $\Lambda _{\rm QCD}=200\;{\rm MeV}, m_b=4.98\;{\rm GeV}$, and $k_+=\Lambda _{\rm QCD}, E_{\gamma} = \frac{m_Q}{4}$, we can show the evaluation of coefficient $C_1^0$ in
Fig.~\ref{Fig5:epsart}.

\section{\label{sec:level7}Numerical Applications}

The amplitude is derived in Eq.~(\ref{eq.7.3}), which can be decomposed as \cite{Sachrajda other work,bsw1}
\begin{equation}
\begin{split}
&<\gamma\mid \bar{q} \Gamma ^{\mu} Q\mid P>=\epsilon _{\mu\nu\rho\sigma}\varepsilon ^{\nu}p_P^{\rho}p_{\gamma}^{\sigma}F_V+i\left(\varepsilon ^{\mu}p_P\cdot p_{\gamma}-p_{\gamma}^{\mu}\varepsilon \cdot p_P\right)F_A\\
\end{split}
\label{eq.7.4}
\end{equation}
The contribution in Fig.~\ref{Fig1:epsart}.c. depends on not only $E_{\gamma}$ but also on $p_{\nu}$ and $p_l$. For simplicity, we treat this term separately, the form factors is written as
\begin{equation}
\begin{split}
&<\gamma\mid \bar{q} \Gamma ^{\mu} Q\mid P>+F_c p_P^{\mu}=\epsilon _{\mu\nu\rho\sigma}\varepsilon ^{\nu}p_P^{\rho}p_{\gamma}^{\sigma}F_V+i\left(\varepsilon ^{\mu}p_P\cdot p_{\gamma}-p_{\gamma}^{\mu}\varepsilon \cdot p_P\right)F_A+F_c p_P^{\mu}\\
&F_V=\frac{1}{(2\pi)^3}\frac{3}{\sqrt{6}}\int d^3 k \Psi (k)\frac{1}{2\sqrt{p_{q0}p_{Q0}(p_{q0}+m_q)(p_{Q0}+m_Q)}} \frac{1}{m_PE_{\gamma}}\\
&\times \left(2e_qC_1m_Q-C_1^0p_{q0}e_q+e_Q 2p_{q0} C_3\right)\\
&F_A=\frac{1}{(2\pi)^3}\frac{3}{\sqrt{6}}\int d^3 k \Psi (k)\frac{1}{2\sqrt{p_{q0}p_{Q0}(p_{q0}+m_q)(p_{Q0}+m_Q)}} \frac{1}{m_PE_{\gamma}}\\
&\times \left(2e_qC_1m_Q-C_1^0p_{q0}e_q-e_q\frac{2p_{q0}m_Q}{E_{\gamma}}C_2-e_Q 2p_{q0} C_3+e_Q\frac{2p_{q0}m_Q}{E_{\gamma}}C_4\right)\\
\end{split}
\label{eq.7.5}
\end{equation}
The relation $F_A = F_V$ at leading order is explicitly broken at order $O(\Lambda _{\rm QCD}\left/m_Q\right.)$ as expected \cite{FaneqFv}. We evaluate this integral using \cite{Teacher Yang wave function}
\begin{equation}
\begin{split}
&m_D=1.9\;{\rm GeV},\;m_B=5.1\;{\rm GeV},\;m_u=m_d=0.08\;{\rm GeV},\;m_b=4.98\;{\rm GeV},\;m_c=1.54\;{\rm GeV}\\
&\Lambda _{\rm QCD}=200\;{\rm MeV},\;\lambda _B = 2.8\;{\rm GeV}^{-1},\;\lambda _D = 3.4\;{\rm GeV}^{-1}%,\lambda _{D_s} = 3.2{\rm GeV}^{-1}
\end{split}
\label{eq.7.6}
\end{equation}

\begin{figure}
\includegraphics[scale=0.8]{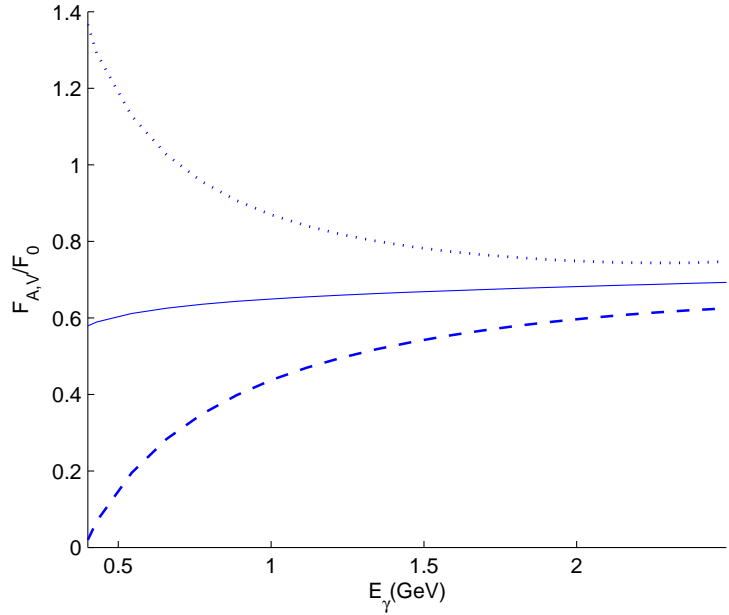}
\caption{\label{Fig6:epsart} Form factors of $B\to \gamma e \nu _e$ as functions of $E_{\gamma}$. The results are presented as the ratios of the form factors at 1-loop order to the one at the tree level and the leading order of the $\Lambda _{\rm QCD}\left/m_Q\right.$ expansion, which is denoted as $F_0$. The solid line is $F_A\left/F_0\right.=F_V\left/F_0\right.$, where $F_{A,V}$ are at the order of $O(\alpha _s(\Lambda _{\rm QCD}\left/m_Q\right.)^0)$. The dotted line is $F_V\left/F_0\right.$ and the dashed line is $F_A\left/F_0\right.$, both at the order of $O(\alpha _s\Lambda _{\rm QCD}\left/m_Q\right.)$.}
\end{figure}

\begin{figure}
\includegraphics[scale=0.75]{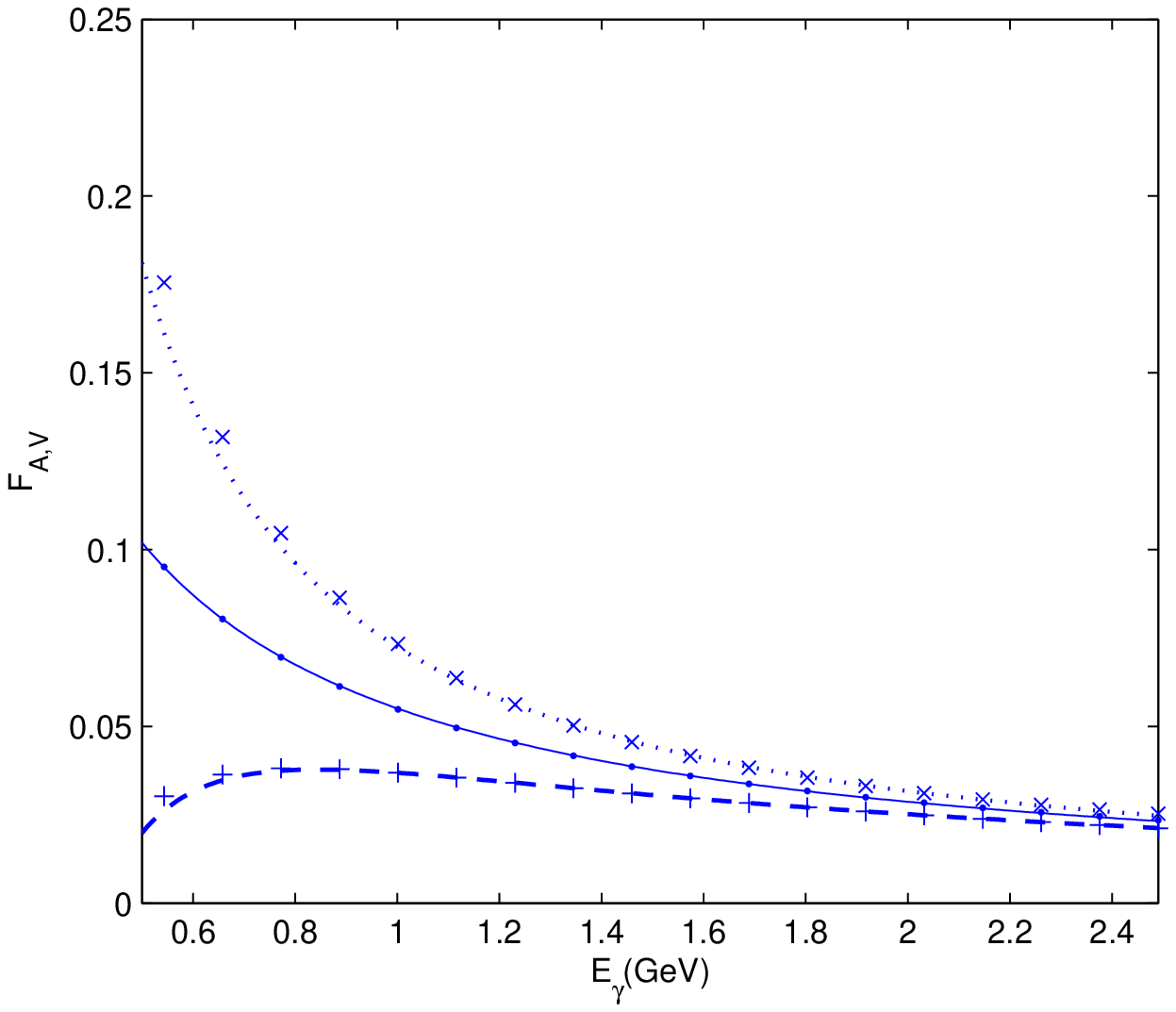}
\caption{\label{Fig7:epsart} Fit of the form factors of $B\to \gamma e \nu _e$. The solid line is for the result at the leading order of $\Lambda _{\rm QCD}\left/m_Q\right.$. The `$\times$' points and the dotted line are for $F_V$ while the dashed line and the `+' points are for $F_A$.}
\end{figure}

\begin{figure}
\includegraphics[scale=0.8]{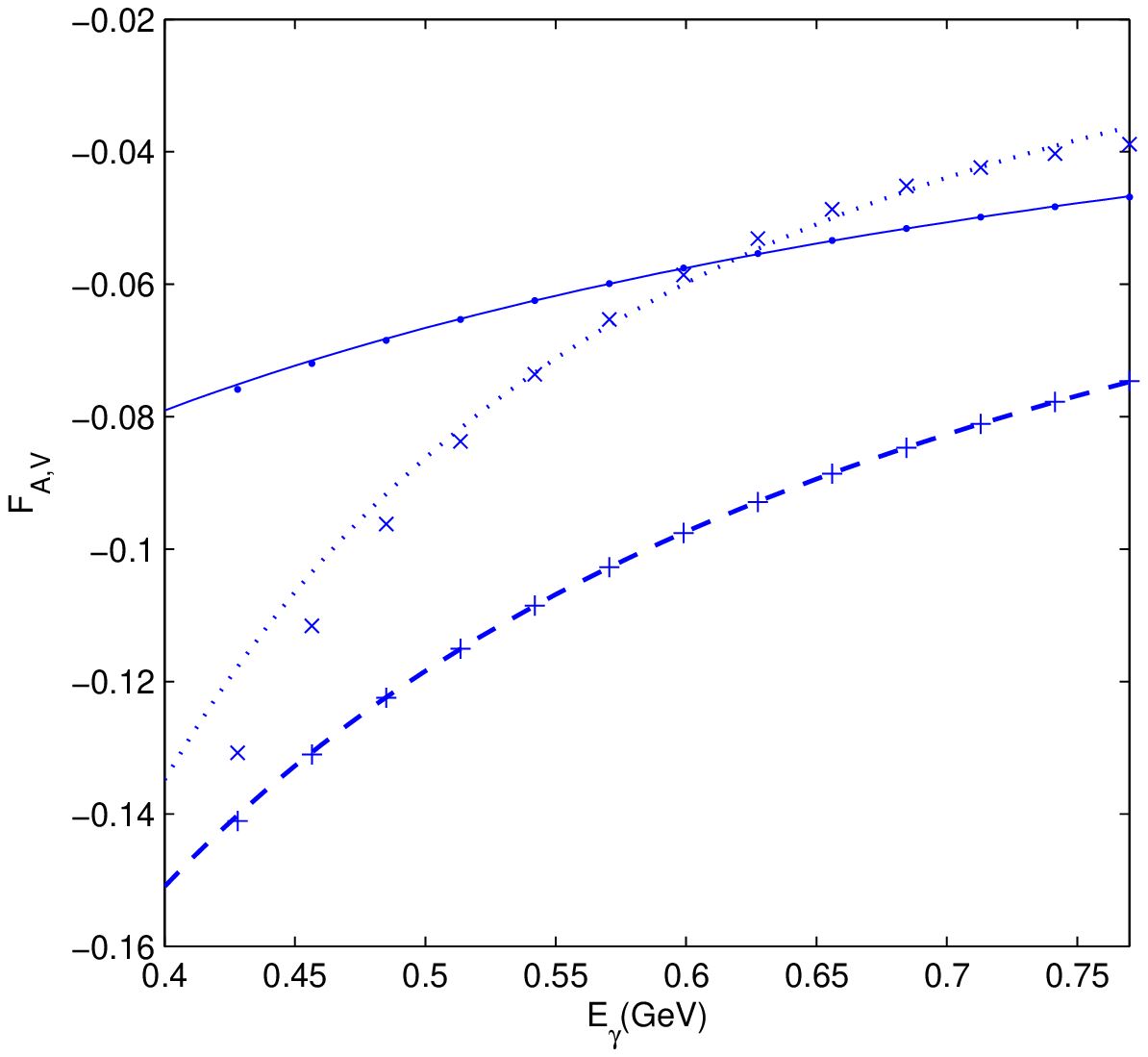}
\caption{\label{Fig9:epsart} Fit of the form factors of $D\to \gamma e \nu _e$. The solid line is for the result at the leading order of $\Lambda _{\rm QCD}\left/m_Q\right.$. The `$\times$' points and the dotted line are for $F_V$ while the dashed line and the `+' points are for $F_A$.}
\end{figure}

The result of $F_{A,V}$ of $B\to \gamma e \nu _e$ is shown in Fig.~\ref{Fig6:epsart}, the $O(\Lambda _{\rm QCD}\left/E_{\gamma}\right.)$ contribution is more important at the region $E_{\gamma}\to 0$ which is clearly shown in the figure. The numerical results of $F_{A,V}$ are inconvenient to use when calculate the decay widths. For simplicity, we use some simple forms to fit the numerical results. For the form factors at the order $O(\alpha _s(\Lambda _{\rm QCD}\left/m_Q\right.)^0)$, we use the single-pole form
\begin{equation}
\begin{split}
&F_A^{\alpha _s}(E_{\gamma})=F_V^{\alpha _s}(E_{\gamma})=\frac{f(0)}{q^2-m^{*2}}=\frac{f(0)}{m_P^2-m^{*2}-2m_PE_{\gamma}}
\end{split}
\label{eq.7.7}
\end{equation}
where $q=p_P-p_{\gamma}$.
While up to the order $O(\alpha _s \Lambda _{\rm QCD}\left/m_Q\right.)$, inspired by the form factors in Ref~\cite{naive factorization} the form factors are fitted as
\begin{equation}
\begin{split}
&F_{A,V}^{\frac{\alpha _s}{m_Q}}(E_{\gamma})=\left(A_{A,V}\frac{\Lambda_{\rm QCD}}{E_{\gamma}}+B_{A,V}\left(\frac{\Lambda_{\rm QCD}}{{E_{\gamma}}}\right)^2\right)
\end{split}
\label{eq.7.8}
\end{equation}
The predicted results for the form factors are more reliable at the region $E_{\gamma}\gg \Lambda _{\rm QCD}$ because we have neglected the higher order terms of $\Lambda _{\rm QCD}\left/E_{\gamma}\right.$ in the numerical calculation, so we choose the region $E_{\gamma}>2\Lambda_{\rm QCD}$ to fit the parameters in Eqs.~(\ref{eq.7.7}) and (\ref{eq.7.8}). The fitting is given in Table.~\ref{table:table1} and shown in Fig.~\ref{Fig7:epsart}. and Fig.~\ref{Fig9:epsart}.
\begin{table}[ht]
\begin{tabular}{ c c c c c c c c }
   & $m^*({\rm GeV})$ & $f(0)({\rm GeV})$ & $A_V$ & $B_V$ & $A_A$ & $B_A$ \\
\hline
  $B\to \gamma e \nu_e$ & $5.37$ & $-0.63$ & $0.27$ & $0.45$ & $0.32$ & $-0.67$ \\
  $D\to \gamma e \nu_e$ & $1.98$ & $-0.15$ & $-0.00095$ & $-0.54$ & $-0.27$ & $-0.059$ \\
\end{tabular}
\caption{The results of the parameters in the form factors in Eqs.~(\ref{eq.7.7}) and (\ref{eq.7.8}).}
\label{table:table1}
\end{table}
On the other hand, $F_c$ can be related to the decay constant \cite{Teacher Yang wave function,naive factorization} by
\begin{equation}
\begin{split}
&F_c p_P^{\mu}=ie\frac{ p_l\cdot \varepsilon P_L^{\mu}}{p_l\cdot p_{\gamma}}<0|\bar{q}\gamma _{\mu}(1-\gamma_5)Q|P>=-ief_Pp_P^{\mu} \frac{ p_l\cdot \varepsilon}{p_l\cdot p_{\gamma}}
\end{split}
\label{eq.7.9}
\end{equation}
Using the fitted result of $F_A$, $F_V$, the result of $F_c$, and using the Cabibbo - Kobayashi - Maskawa (CKM) matrix elements \cite{particaldatagroup,vub}
\begin{equation}
\begin{split}
&V_{\rm cd}=0.226,\;V_{\rm ub}=0.0047\\
\end{split}
\label{eq.7.10}
\end{equation}
we obtain the result for the branching ratios. There are IR divergences in the radiative leptonic decays in the case that the photon is soft or the photon is collinear with the emitted lepton. Theoretically this IR divergences can be canceled by adding the decay rate of the radiative leptonic decay with the pure leptonic decay rate, in which one-loop correction is included \cite{changch}. The radiative leptonic decay can not be distinguished from the pure leptonic decay in experiment when the photon energy is smaller than the experimental resolution to the photon energy. So the decay rate of the radiative leptonic decay depend on the experimental resolution to the photon energy $E_{\gamma}$ which is denoted by $\Delta E_{\gamma}$. The dependence of the branching ratios on the resolution are listed in Table.~\ref{table:table2}.
\begin{table}[ht]
\begin{tabular}{ c c c| c c c }
\hline
    $\Delta E_{\gamma}$ & $BR(B\to e\nu _e \gamma)$ & $BR(D\to e\nu _e \gamma)$ & $\Delta E_{\gamma}$ & $BR(B\to e\nu _e \gamma)$ & $BR(D\to e\nu _e \gamma)$\\
\hline
  $5{\rm MeV}$ & $1.77\times 10^{-6}$ & $3.10\times 10^{-5}$ & $20{\rm MeV}$ & $1.56\times 10^{-6}$ & $2.53\times 10^{-5}$\\
  $10{\rm MeV}$ & $1.66\times 10^{-6}$ & $2.81\times 10^{-5}$ & $25{\rm MeV}$ & $1.53\times 10^{-6}$ & $2.45\times 10^{-5}$\\
  $15{\rm MeV}$ & $1.60\times 10^{-6}$ & $2.64\times 10^{-5}$ & $30{\rm MeV}$ & $1.48\times 10^{-6}$ & $2.38\times 10^{-5}$\\
  \hline
\end{tabular}
\caption{The branching ratios with different photon resolution $\Delta E_{\gamma}$.}
\label{table:table2}
\end{table}

\begin{table}[ht]
\begin{tabular}{ c c c c }
   & $BR_{O( (\frac{\alpha _s\Lambda _{\rm QCD}}{m_Q})^0)}$ & $BR_{O(\alpha _s(\frac{\Lambda _{\rm QCD}}{m_Q})^0)}$ & $BR_{O(\frac{\alpha _s \Lambda _{\rm QCD}}{m_Q})}$\\
\hline
  $B\to e\nu _e \gamma$ & $2.38\times 10^{-6}$ & $1.01\times 10^{-6}$ & $1.66\times 10^{-6}$ \\
%  $D_s\to e\nu _e \gamma$ & $6.11\times 10^{-5}$ & $1.30\times 10^{-5}$ & $5.13\times 10^{-4}$ \\
  $D\to e\nu _e \gamma$ & $9.62\times 10^{-6}$ & $2.71\times 10^{-6}$ & $2.81\times 10^{-5}$ \\
\end{tabular}
\caption{The branching ratios of the decay modes.}
\label{table:table3}
\end{table}

Using $\Delta E_{\gamma} = 10{\rm MeV}$ \cite{photon res}, the branching ratios are given in Table.~\ref{table:table3}.
We find that, in general, the 1-loop results are smaller than the tree level results. The 1-loop correction is found to be important. For B meson, the correction to the decay amplitude at the order $O(\Lambda _{\rm QCD}\left/m_Q\right.)^0$ is about $10\%$ to $30\%$ due to the large logarithms.

The contribution of the order $O(\alpha _s \Lambda _{\rm QCD}\left/m_Q\right.)$ are generally not negligible. For $B$ meson, the correction of the order $O(\Lambda _{\rm QCD}\left/m_Q\right.)$ contribution to the decay amplitude can be as large as $30\%$. For $D$ mesons, the mass of $c$ quark is not large enough, the order $O(\Lambda _{\rm QCD}\left/m_Q\right.))$ contributions is much more important, it is necessary to include higher order corrections in $\Lambda _{\rm QCD} \left/ m_Q\right.$ expansion.

\section{\label{sec:level8}Conclusion}

In this paper, we study the factorization of the radiative leptonic decays of $B^-$ and $D^-$ mesons. Compared with the work in Ref.~\cite{Genon and Sachrajda Radivative Leptonic}, the factorization is extended to include the $O(\Lambda _{\rm QCD}\left/m_Q\right.)$ contributions, and the transverse momentum is also considered. The factorization is proved explicitly at 1-loop order, the valid region of the factorization is extended. The hard kernel at order $O(\alpha _s \Lambda _{\rm QCD}\left/m_Q\right.)$ is obtained. We use the wave function obtained in Ref.~\cite{Teacher Yang wave function} to derive the numerical results. The branching ratios of $B^-\to \gamma e\bar{\nu}$ is found to be at the order of $10^{-6}$, which is close to the previous works \cite{naive factorization,TM Yan work,prework1,prework2,prework3}. In the previous works, the results of $D$ mesons are different from each other from $10^{-3}$ to $10^{-6}$, our results agree with $10^{-5}$.

%What is the main difference
We also find that the $O(\Lambda _{\rm QCD}\left/m_Q\right.)$ contribution is very important even for $B$ meson, the correction to the decay amplitude is about $20\% - 30\%$, which can affect the branching ratios about $50\%$. This is because of the importance of $O(\Lambda _{\rm QCD}\left/E_{\gamma}\right.)$ contributions. In previous works, $O(\Lambda _{\rm QCD}\left/E_{\gamma}\right.)$ contributions is neglected. For a typical region, $E_{\gamma}\sim m_Q\left/4\right.$, which is also the leading region of the phase space of the tree level, the neglected contributions can be up to $20\%$. As a result, the correction to the branching ratios can be up to $40\%$ at tree level.

\textbf{Acknowledgements}: This work is supported in part by the
National Natural Science Foundation of China under contracts Nos.
11375088, 10975077, 10735080, 11125525.

\begin{appendices}
\renewcommand{\thesection}{Appendix \Alph{section}}

\begin{center}
\section{\label{sec:ap1}IBP reductant relation of wave function}
\end{center}

The integral is
\renewcommand{\theequation}{\Alph{section}. 1}
\begin{equation}
\begin{split}
&\Phi_q^{(1)}(k_{\bar{q}}, k_Q)=\int d^4x \int d^4y e^{ik_{\bar{q}}\cdot x}e^{ik_Q\cdot y}<0|\bar{q}_{\bar{q}}(x)ig_s\int _y^x dz z_{\mu}A^{\mu}(z) Q(y)\\
&\times ig_s\int d^4x_2 \bar{q} _{\bar{q}}(x_2)\slashed {A}(x_2)q _{\bar{q}}(x_2)|\bar{q}^S(p_{\bar{q}}),Q^s(p_Q)>
\end{split}
\label{eq.a.1}
\end{equation}
After variable substitution using $z=x+\alpha y$, we find
\renewcommand{\theequation}{\Alph{section}. 2}
\begin{equation}
\begin{split}
\int _y^x dz z_{\mu}A^{\mu}(z)&=\int _0^1 d (y + \alpha (x-y))_{\mu}A^{\mu}(y + \alpha (x-y))=\int _0^1 d\alpha (x-y)_{\mu}A^{\mu}(y + \alpha (x-y))
\end{split}
\label{eq.a.2}
\end{equation}
After the contraction and then integral over $x_2$, $p$, the result is
\renewcommand{\theequation}{\Alph{section}. 3}
\begin{equation}
\begin{split}
&\Phi_q ^{(1)}\otimes T^{(0)}=-C_Fg_s^2\int d^4x \int d^4y\int _0^1 d\alpha \int \frac{d^d l}{(2\pi)^d}\\
&\times e^{iy\dot(k_Q+\alpha l-p_Q)}e^{ix\dot(k_{\bar{q}}-\alpha l-p_{\bar{q}})}\bar{v}_{\bar{q}}\gamma \cdot(x-y)\frac{1}{l^2}\frac{(-\slashed l-\slashed p_{\bar{q}} + m_{\bar{q}})}{(l+p_{\bar{q}})^2-m_{\bar{q}}^2}T^{(0)}u_Q\\
\end{split}
\label{eq.a.3}
\end{equation}
The integral over $x$ and $k_{\bar{q}}$ can be worked out term by term. However, we find IBP reduction relation \cite{IBP} an elegant way to do so. Consider this integral
\renewcommand{\theequation}{\Alph{section}. 4}
\begin{equation}
\begin{split}
&\int \frac{d^4k_q}{(2\pi)^4}\Gamma(k_q)\int d^4xe^{-i(p_{\bar{q}}+\alpha l-k_{\bar{q}})\cdot x}
\end{split}
\label{eq.a.4}
\end{equation}
It is unchanged when $k_q$ is shifted, so under the infinitesimal transformation
\renewcommand{\theequation}{\Alph{section}. 5}
\begin{equation}
\begin{split}
&k_q\to k_q+\beta K\\
\end{split}
\label{eq.a.5}
\end{equation}
The integral transforms as
\renewcommand{\theequation}{\Alph{section}. 6}
\begin{equation}
\begin{split}
\Gamma(k_q)\int d^4x e^{-i(p_{\bar{q}}+\alpha l-k_{\bar{q}})\cdot x}\to \left(\beta K\cdot \frac{\partial}{\partial k_q}\right)\left(\Gamma(k_q)\int d^4x e^{-i(p_{\bar{q}}+\alpha l-k_{\bar{q}})\cdot x}\right)
\end{split}
\label{eq.a.6}
\end{equation}
The Lie algebra leads to
\renewcommand{\theequation}{\Alph{section}. 7}
\begin{equation}
\begin{split}
&\int \frac{d^4k_q}{(2\pi)^4}\left(K\cdot \frac{\partial}{\partial k_q}\right)\left(\Gamma(k_q)\int d^4x e^{-i(p_{\bar{q}}+\alpha l-k_{\bar{q}})\cdot x}\right)=0\\
\end{split}
\label{eq.a.7}
\end{equation}
so that
\renewcommand{\theequation}{\Alph{section}. 8}
\begin{equation}
\begin{split}
&\int \frac{d^4k_q}{(2\pi)^4} \Gamma(k_q) K \cdot x\int d^4x e^{-i(p_{\bar{q}}+\alpha l-k_{\bar{q}})\cdot x}=i \left.K\cdot \frac{\partial \Gamma(k_q)}{\partial k_q}\right|_{k_q=p_{\bar{q}}+\alpha l}
\end{split}
\label{eq.a.8}
\end{equation}
So after integrate over $x$, $k_{\bar{q}}$, $y$ and $k_Q$, the result is
\renewcommand{\theequation}{\Alph{section}. 9}
\begin{equation}
\begin{split}
&\Phi_q ^{(1)}\otimes T^{(0)}=ig_s^2C_F\int \frac{d^d l}{(2\pi)^d} \frac{1}{l^2}\bar{v}_{\bar{q}}\gamma ^{\rho}\frac{(\slashed {l}+\slashed {p}_{\bar{q}} - m_{\bar{q}})}{(l+p_{\bar{q}})^2-m_{\bar{q}}^2} \int _0^1 d\alpha\left.\left(\frac{\partial T^{(0)}}{\partial k_q^{\rho}}-\frac{\partial T^{(0)}}{\partial k_Q^{\rho}}\right)\right| _{k_q=k',k_Q=K'} u_Q\\
&k'=p_{\bar{q}}+\alpha l,\;\;K'=p_Q-\alpha l\\
\end{split}
\label{eq.a.9}
\end{equation}

\end{appendices}

\end{document}